\documentclass[11pt,a4paper]{article}
\pdfoutput=1
\usepackage[pdftex]{graphics}
\usepackage{jheppub}
\usepackage{amsmath,amssymb,amsfonts}
\usepackage{array,booktabs}
\usepackage{slashed}

\newcommand{\be}{\begin{eqnarray}}
\newcommand{\ee}{\end{eqnarray}}

\newcommand{\bn}{\begin{enumerate}}
\newcommand{\en}{\end{enumerate}}
\newcommand{\bl}{\begin{align}}
\newcommand{\el}{\end{align}}

\def\a{\alpha}
\def\b{\beta}
\def\g{\gamma}
\def\dt{\delta}

\def\e{\epsilon}
\def\ve{\varepsilon}

\def\l{\lambda}
\def\m{\mu}
\def\n{\nu}

\def\r{\rho}

\def\s{\sigma}

\def\L{\Lambda}

\def\O{\Omega}

\def\det{{\rm det}}

\def\da{{\dot{\a}}}
\def\db{{\dot{\b}}}
\def\dg{{\dot{\g}}}
\def\dd{{\dot{\d}}}

\def\jmath{{j}}

\def\deq{{\;\; \dot{=} \;\;}}

\def\c{{\gamma}}
\def\dc{{\dot{\c}}}
\def\d{{\delta}}
\def\be{{\bar{\epsilon}}}
\def\bs{{\bar{\sigma}}}

\def\mem{\hspace{0.1em}}

\def\Re{{\operatorname{Re}}}
\def\Im{{\operatorname{Im}}}

\def\rambda{{\bar{\lambda}}}
\def\lmu{{\bar{\mu}}}   \def\rmu{{\mu}}  

\def\rmA{{\mathrm{A}}}

\usepackage[dvipsnames]{xcolor}
\definecolor{purple}{RGB}{171,1,207}
\definecolor{green}{rgb}{0.10, 0.57, 0.19}

\title{The Relativistic Spherical Top as a Massive Twistor} 

\author[a]{Joon-Hwi Kim}
\author[b]{Jung-Wook Kim}
\author[a,c,d]{Sangmin Lee} 

\affiliation[a]{Department of Physics and Astronomy, Seoul National University, Seoul 08826, Korea}
\affiliation[b]{Centre for Research in String Theory, School of Physics and Astronomy,
\\
Queen Mary University of London, Mile End Road, London E1 4NS, United Kingdom}  
\affiliation[c]{Center for Theoretical Physics, Seoul National University, Seoul 08826, Korea}
\affiliation[d]{College of Liberal Studies, Seoul National University, Seoul 08826, Korea}

\abstract{
We prove the equivalence between two traditional approaches 
to the classical mechanics of a massive spinning particle in special relativity.
One is the spherical top model of Hanson and Regge, recast in a Hamiltonian formulation 
with improved treatment of covariant spin constraints. 
The other is the massive twistor model, slightly generalized to incorporate the Regge trajectory relating the mass to the total spin angular momentum. 
We establish the equivalence by computing the Dirac brackets of the physical phase space carrying three translation and three rotation degrees of freedom.
Lorentz covariance and little group covariance 
uniquely determine the structure of the physical phase space.   
The Regge trajectory does not affect the phase space but enters the equations of motion. 
Upon quantization, the twistor model produces a spectrum that agrees perfectly with the massive spinor-helicity description 
proposed by Arkani-Hamed, Huang and Huang for scattering amplitudes for all masses and spins.
}

\emailAdd{joonhwi.kim@snu.ac.kr, jung-wook.kim@qmul.ac.uk, sangmin@snu.ac.kr}

\begin{document}
\begin{flushright}
\vspace{10pt} \hfill{QMUL-PH-21-08} \vspace{20mm}
\end{flushright}
\maketitle

\section{Introduction}

The Lagrangian for a free massive particle in special relativity is a rare gem; being simpler than its 
Newtonian predecessor, it cries out for a leap to general relativity. 
Unfortunately, this extreme simplicity is quickly lost as soon as the particle starts spinning, even if the particle remains free.
All textbooks explain that the quantum mechanics of an elementary particle of minimal nonzero spin is governed by the Dirac equation. 
As for the classical mechanics of a relativistic particle with arbitrary (possibly macroscopic) spin, 
there seems to be no such consensus.

A Lorentz covariant description of a spinning particle inevitably introduces constraints. 
The mass-shell constraint ($p^2 + m^2=0$) reveals that only three of 
the four translation degrees of freedom are independent. 
Identifying and imposing the constraints for the rotation degrees of freedom is a major source of complication 
in any model of a spinning particle. 
Different choices are made with not-so-obvious advantages or disadvantages. 

To simplify the discussion somewhat, we restrict our attention 
to the relativistic analog of the ``spherical top"; 
in the rest frame of the particle, the model should retain the SO$(3)$ rotation symmetry. 
By a slight abuse of language, 
we refer to this SO$(3)$ as the ``little group".
Over many decades, a variety of models for a massive spinning particle have been proposed. It is not always straightforward to
tell whether two models are equivalent or not; 
see {\it e.g.}  
\cite{Frydryszak:1996mu,Rivas:2002,Deriglazov:2017jub} for reviews with a comprehensive bibliography. 

From a Hamiltonian point of view, ignoring the translation temporarily and focusing on the rotation, 
we can distinguish three types of phase spaces with a manifest SO$(3)$ isometry: 
$S^2$, $T^*(S^2)$, $T^*(\mathrm{SO}(3))$, where $T^*(M)$ denotes the cotangent bundle of a manifold $M$. 
But, even if two models appear to have phase spaces of different dimensions, 
there remains a possibility that imposing an extra constraint on one model reproduces the other. 

The main goal of this paper is to reduce the catalog of 
competing models to point out a hitherto unnoticed equivalence 
between two famous models of a massive spinning particle.
One of them is the spherical top model of Hanson and Regge \cite{Hanson:1974qy}. It uses the Lorentz covariant generalization of the Euler angles. 
Before constraints are imposed, the rotation degrees of freedom 
are described by an element of the Lorentz group SO$(1,3)$ such that the spin angular momentum becomes an anti-symmetric tensor with six components. 
The other model is the massive twistor (a.k.a.  two-twistor or bi-twistor) model. 
The dynamical variables consist of a pair of usual twistors, amounting to eight complex variables in total. 

We should clarify what we mean by the two models at the technical level.  
Hanson and Regge \cite{Hanson:1974qy} began with a Lagrangian formulation and moved on to a constrained Hamiltonian system. 
They showed that the mass can be an arbitrary function of the total 
spin angular momentum. We will call the function ``Regge trajectory".
For a systematic analysis of constraints, it is more convenient to start directly from a Hamiltonian formulation, 
where the choice of Regge trajectory replaces the choice of Lagrangian in the original formulation.   
We begin with a recent proposal for constraints for spins by Steinhoff \cite{Steinhoff:2015ksa} and gradually deviate from it. 
On the twistor side, while the key ideas for a massive twistor model had been available since the beginning of 
the twistor theory \cite{Penrose:1972ia,Newman:1974fr,Tod:1976sk,Hughston:1979pg,Huggett:1986fs}, 
a particular version 
of the model suitable for our purposes was constructed rather recently by Fedoruk and Lukierski~\cite{Fedoruk:2014vqa}.
They considered a two-twistor generalization of Shirafuji's twistor model for massless spinning particles \cite{Shirafuji:1983zd} but specialized to the case of constant mass. 
We generalize their model slightly further by incorporating the Regge trajectory.

Having specified the two models, we give an overview of how to prove the equivalence.
Before embarking on a technical discussion, 
let us count the dimensions of the physical phase spaces. 
In the spherical top model, among the six rotation degrees of freedom, three (loosely corresponding to boosts) are removed by constraints. Thus, the physical phase space becomes 12 dimensional (three coordinate-momenta pairs for translation and 
three pairs for rotation). 
In the massive twistor model, 
two complex-valued functions of the twistor variables are set to zero, leaving behind six complex (12 real) variables.

Choosing and imposing constraints in the two models is by no means a new subject. Our choices differ from previous ones in a few aspects. 
In the spherical top model, we begin with a slight variation of the ``spin-gauge" generators of Steinhoff \cite{Steinhoff:2015ksa}. 
The three generators form first-class constraints. 
We pair them with three ``gauge-fixing conditions" 
to form a set of second-class constraints. 
Including the mass-shell constraints and the corresponding gauge-fixing condition, we have four pairs of constraints in total.
It is straightforward to compute the Dirac brackets taking account of all constraints, which reveal the physical phase space 
without leaving the original coordinates.

The massive twistor space is equipped with a large symmetry: 
$\mathrm{SU}(2,2)\times \mathrm{SU}(2)\times \mathrm{U}(1)$. 
The SU$(2,2)$ is identified with the SO$(2,4)$ conformal symmetry of the Minkowski space, 
to be broken to the Lorentz group by the mass of the particle. 
The SU$(2)$ is to be identified with the SO$(3)$ little group. 
To emphasize this point, following \cite{Fedoruk:2014vqa} and differing from some earlier works, 
we use a manifestly SU$(2)$ covariant notation throughout. 
The U$(1)$ may be associated with electric charge \cite{Newman:1974fr}, 
but we focus on neutral particles only. 
Again, the constraints come in pairs: gauge generators and gauge-fixing conditions. 
The gauge generators are twistor counterpart of the mass-shell constraints. 
The real part of the gauge-fixing conditions corresponds to the dilatation symmetry 
and the imaginary part to the internal U$(1)$ symmetry. 
Computing the Dirac brackets on the constrained twistor space 
is straightforward. 

To establish the equivalence between the two models, a map between the two phase spaces are in order. 
We propose and verify the map between the two models in two equivalent ways. 
One is to compare the Dirac brackets and the other is 
to compare the invariant coordinates that solve the constraints in the respective models. 
The map is determined, to a large extent, by Lorentz symmetry and little group symmetry. 
Remarkably, the 12-dimensional physical phase space turns out to be a copy of $T^*(\mathrm{SO}(1,3))$ 
and is insensitive to the Regge trajectory. 
The Regge trajectory contributes to the dynamics only through the equations of motion. 

Section~\ref{sec:vect} on the spherical top model and section~\ref{sec:spin} on the massive twistor model 
constitute the main body of this paper. 
In section~\ref{sec:vect}, we review and slightly modify Steinhoff's proposal for spin-gauge generators and pair them with covariant gauge-fixing conditions. 
After giving an explicit description of the gauge orbits, we analyze the fully constrained physical phase space. 
We apply two equivalent methods. One is the Dirac bracket in the original coordinates, 
the other is to find new coordinates for the physical phase space. 
We also use the constraints to derive the covariant equations of motion and confirm that the solutions indeed describe a free spinning particle. 
In section~\ref{sec:spin}, we take parallel steps 
in the massive twistor model. 
Although we are not the first to construct the model, 
we motivate each step in a self-contained manner 
in such a way that the map to the spherical top model is forced upon us 
under mild assumptions. 
We spell out the constraints, compute the Dirac brackets, and derive the equations of motion. 
The equivalence with the spherical top model is established quite explicitly. 
Upon quantization, the spectrum of the massive twistor model coincides precisely 
with the spectrum of massive states expressed in terms of the ``massive helicity-spinors" \cite{Arkani-Hamed:2017jhn} in the context of 
scattering amplitudes of quantum fields. 
Section~\ref{sec:discussion} discusses a few possible applications of our results. 

\section{Spherical top} \label{sec:vect}

We revisit the celebrated spherical top model by Hanson and Regge \cite{Hanson:1974qy} from a Hamiltonian perspective, 
taking hints from a recent  work by Steinhoff \cite{Steinhoff:2015ksa}.

\subsection{Spin-gauge symmetry}

The spherical top model has manifest Lorentz symmetry. 
Its translation is described by the position 4-vector $x^\mu$, 
and its orientation is described by an $\mathrm{SO}(1,3)$ matrix $\L^\m{}_A$ 
which defines a ``body-attached orthonormal frame'':
\begin{align}
    \eta_{\m\n} \L^\m{}_A \L^\n{}_B = \eta_{AB} \,,\qquad \L^\m{}_A \L^\n{}_B \eta^{AB} = \eta^{\m\n} \,,
    \label{BAOF}
\end{align}
where the Greek and uppercase Latin indices lie in the range $\m, A = 0, \cdots, 3$. 
We work in the mostly plus signature. 
The canonical momenta of $x^\m$ and $\Lambda^\m{}_A$ are denoted by $p_\mu$ and $S_{\m\n}$, respectively. 
The non-vanishing Poisson brackets on the phase space satisfying \eqref{BAOF} are
\begin{align}
\begin{split}
\{ x^\mu , p_\nu \} &= \delta^\mu{}_\nu \,, 
\\
\{ \L^\r{}_A, S_{\mu\nu} \} &= -( \delta^\r{}_\m  \Lambda_{\n A} - \delta^\r{}_\n \Lambda_{\m A} )
\,,
\\
\{ S_{\mu\nu} , S_{\rho\sigma} \} &= 
- (\eta_{\nu\rho} S_{\mu \sigma} - \eta_{\mu\rho} S_{\nu\sigma} 
- \eta_{\nu\sigma} S_{\mu\rho} + \eta_{\mu\sigma}S_{\nu\rho} ) 
\,.
\label{poisson-1}
\end{split}
\end{align}
The spin satisfies the same relation as the orbital angular momentum $L_{\m\n} := x_\m p_\n - x_\n p_\m$. 
We denote the total angular momentum by $J_{\m\n} = L_{\m\n} + S_{\m\n}$.
The Poisson bracket is equivalent to the symplectic form, 
\begin{align}
\begin{split}
\omega  = d\theta &= dp_\m \wedge dx^\m + \frac{1}{2} dS_{\m\n} \wedge \Theta^{\m\n} - \frac{1}{2} S_{\m\n}\mem  \Theta^{\m\r} \wedge \Theta_\r{}^\n
\,,
\\
\theta &= p_\mu dx^\mu + \frac{1}{2} S_{\m\n}\Theta^{\m\n}  \,,
\qquad \Theta^{\m\n} = \L^{\m A} d\L^\n{}_A \,.
\label{symp-1}
\end{split}
\end{align}
The symplectic form appears to be invariant under two Lorentz symmetries, 
one acting on the space-time index $\m$ and 
the other acting on the body frame index $A$. 
As we will see shortly, the body frame symmetry is broken to the little group SO$(3)$, 
so that the time-like ($A=0$) component and the space-like ($A=a=1,2,3$) components 
do not have to be correlated. 
Nevertheless, as a bookkeeping device, 
we will often combine a little group scalar and a little group vector 
as if they form a ``4-vector.''

The Hamiltonian mechanics of the spherical top model is nontrivial because the phase space coordinates in \eqref{poisson-1} are constrained. 
Even for a spin-less particle, incorporating the mass-shell constraint, 
\begin{align}
p^2 + m^2 = 0 
\quad 
(p^2 := p^\m p_\m) \,,
\end{align}
in a manifestly covariant way requires some thoughts. Constraining the rotation degrees of freedom requires extra effort. 
Intuitively, among the six apparent degrees of freedom associated with $(\Lambda_\m{}^A , S_{\m\n})$, the ``boost-like" three should be removed 
and only the three genuine rotation degrees of freedom should be retained. 

Our discussion of the spin constraints begins with a recent proposal by Steinhoff \cite{Steinhoff:2015ksa} 
who introduced the spin-gauge generators:
\footnote{
This choice is closely related to the Pryce-Newton-Wigner choice \cite{Pryce:1948pf,Newton:1949cq}. 
A nearly identical set of first-class constraints appeared earlier in \cite{Nielsen:1995cy}.
Refs.~\cite{Levi:2015msa,Levi:2018nxp} discuss the importance of spin-gauge symmetry in effective field theories of post-Newtonian gravity.
}
\begin{align}
\mathcal{C}_\mu =  (\hat{p}^\rho + \Lambda^\rho{}_0) S_{\rho\mu}\,, 
\quad 
\hat{p}^\rho = p^\rho/|p| \,, 
\quad 
|p| = \sqrt{-p^2}\,.
\label{eq:spin-gauge-stein}
\end{align}
They form a set of first-class constraints:
\begin{align}
\{ \mathcal{C}_\mu , \mathcal{C}_\nu \} 
=   \mathcal{C}_\m \hat{p}_\n - \mathcal{C}_\n \hat{p}_\m  \,.
\end{align}
As such, they qualify as generators of the spin-gauge symmetry 
in the standard nomenclature for constrained Hamiltonian systems~\cite{Dirac:text,Henneaux:1992ig}; see appendix~\ref{sec:constrained} for a quick review.

Our first step away from Steinhoff's proposal is to rewrite the constraints adapted to the body frame:
\begin{align}
\phi_a := \frac{1}{2}(\hat{p}^\mu + \Lambda^\m{}_0) S_{\mu\nu} \Lambda^\n{}_a  
\quad (a=1,2,3) \,.
\end{align}
This rewriting makes it clear that only three components of $\mathcal{C}_\m$ are independent. Another advantage of the rewriting is that the three generators commute under the Poisson bracket:
\begin{align}
\{ \phi_a , \phi_b \} = 0 \,.
\end{align}
In other words, the gauge algebra is abelian. 
As emphasized by Steinhoff, 
various spin supplementary conditions can be treated in a uniform manner as different gauge-fixing conditions for the same spin-gauge symmetry.
We will focus on the covariant gauge in the rest of this paper. 
The implications of the covariant gauge include 
\begin{align}
\mbox{both} \quad 
\L^\m{}_0 = \hat{p}^\m \quad \mbox{and} \quad S_{\m\n} p^\n = 0 \,. 
\label{cov-gauge-old}
\end{align}
An attempt to use $S_{\m\n}p^\n = 0$ as first-class constraints as an alternative to \eqref{eq:spin-gauge-stein} may cause difficulties discussed in {\it e.g.} 
\cite{Deriglazov:2015bqa}. 
We circumvent the difficulty 
by pairing the spin-gauge generators \eqref{eq:spin-gauge-stein} 
with the following ``canonically conjugate" gauge-fixing conditions,   
\begin{align}
\begin{split}
\chi^a =  \hat{p}_\m \Lambda^{\m a}   \,.
\end{split}
\label{def:cov-gauge-fixing}
\end{align}
Clearly, imposing both $\phi_a = 0$ and $\chi^a = 0$ (and a choice of sign) implies \eqref{cov-gauge-old}. 

Sometimes, it is useful to introduce a conjugate constraint for the mass-shell constraint. 
All together, we have four pairs of constraints: 
\begin{align}
\begin{split}
&\phi_0 = \frac{1}{2}(p^2 + m^2) \,, \quad 
\phi_a = \frac{1}{2}(\hat{p}^\mu + \Lambda^\mu{}_0) S_{\mu\nu} \Lambda^\n{}_a \,,
\\
&\chi^0 = \frac{1}{p^2} x^\mu p_\mu \,, \qquad \quad 
\chi^a =  \hat{p}_\mu \Lambda^{\m a} \,.
\end{split}
\label{constraints-vect-all}
\end{align}
We have adjusted the constant coefficients such that 
\begin{align}
\{\chi^0, \phi_0 \} \approx 1 \,,
\quad 
\{ \chi^a , \phi_b \} \approx \delta^a{}_b \,,
\end{align}
with all the other Poisson brackets weakly vanishing. 
The weak equality $(\approx)$ means that the equality holds on the constraint surface $\phi_A = 0 = \chi^A $. 
In the rest of this paper, we will not explicitly distinguish 
the weak equality from the strong equality, hoping 
that the context will clearly indicate which one is being used.

\paragraph{Regge trajectory} 

Following Steinhoff, we turn on the Regge trajectory 
using the spin-gauge  invariant combination of the spin tensor, 
\begin{align}
m = m(\tilde{S}^2) \,,
\quad 
\tilde{S}^2 := \frac{1}{2} \tilde{S}^{\m\n} \tilde{S}_{\m\n} 
\,,
\quad 
\tilde{S}_{\m\n} := \hat{\eta}_{\m\r} S^{\r\s} \hat{\eta}_{\s\n} \,, 
\quad 
\hat{\eta}_{\mu\nu} := \eta_{\m\n} + \hat{p}_\m \hat{p}_\n \,.
\label{Regge-t}
\end{align}
It is easy to verify that $\{ \phi_a , \tilde{S}_{\m\n} \} = 0$, 
which implies that $\{ \phi_0 , \phi_a \} = 0$ even after the Regge trajectory is turned on. 
Physically, $\tilde{S}_{\m\n}$ is the projection of $S_{\m\n}$ onto the space-like three dimensions transverse to the momentum vector. Compared with the normalized Pauli-Lubanski vector 
$W_\m = - \frac{1}{2} \ve_{\m\n\r\s} J^{\n\r} \hat{p}^\s$ $(\varepsilon_{0123}=+1)$, we have $W^2 = \tilde{S}^2$.

\subsection{Gauge orbit in the phase space} \label{sec:orbit}

The four conjugate pairs of constraints \eqref{constraints-vect-all} 
define the gauge orbits in the unconstrained phase space. 
The dynamics of the spherical top depends only on  
the physical phase space defined by our choice of gauge slice: $\phi_A = 0 = \chi^A$.  

In this subsection, we study the geometry of the gauge orbits. 
We focus on the spin-gauge symmetry $(\phi_a, \chi^b)$ here 
and leave $(\phi_0,\chi^0)$ for a later subsection, 
since the latter pair governs the time-evolution of the spherical top.

\paragraph{Flow equations}

Recall our definition of spin-gauge generators:
\begin{align}
    \phi_a = \frac{1}{2}(\hat{p}^\mu + \Lambda^\mu{}_0) S_{\mu\nu} \Lambda^\nu{}_a \,.
    \nonumber 
\end{align}
We showed earlier that the algebra is abelian, so the gauge orbit should look like $\mathbb{R}^3$ locally. 
How is the gauge orbit embedded in the unconstrained phase space?
 
Let $y^a$ be the coordinates for the $\mathbb{R}^3$. 
With hindsight, we rescale the coordinates slightly such that the flow equations 
for the gauge orbits are given by
\begin{align}
\frac{\partial X}{\partial y^a} = 2 \{ X , \phi_a \} \,,
\end{align}
where $X$ is any one of the phase space coordinates. 
Explicitly, 
\begin{align}
    \begin{split}
    \partial_a p_\mu &= 0 \,,
    \\
    \partial_a \L^\m{}_B &=  \L^\mu{}_a [ (\hat{p} + \Lambda_0)^\rho \Lambda_{\rho B}] - (\hat{p} + \Lambda_0)^\mu \eta_{a B} \,,
    \\
    \partial_a S_{\m\n} &=   (\hat{p}_\mu S_{\nu\rho} - \hat{p}_\nu S_{\mu\rho}  ) \L^\rho{}_a \,,
    \\
    \partial_a x_\mu &=  \frac{1}{|p|} \hat{\eta}_{\mu\nu} S^\nu{}_\rho  \L^\rho{}_a \,.
    \end{split}
\end{align}
To set the initial condition for the equation, 
we demand the covariant gauge conditions $(\phi_a = 0 = \chi^a)$ at the origin of the $y$-space. 
As we move away from the origin, $\phi_a$ stay zero but $\chi^a$ develop non-zero values. 
Since $\{\chi^a , \phi_b\} = \delta^a{}_b$ at the origin, for a small departure from the origin, 
$\chi^a = 2 y^a + \mathcal{O}(y^2)$ should hold. 

\paragraph{Spin-gauge invariants}

Steinhoff \cite{Steinhoff:2015ksa} suggests a number of gauge  invariant combinations that are conserved along the flow. 
In addition to $p_\m$ and $\phi_a$, the list of invariants includes 
\begin{align}
    \begin{split}
        &\tilde{\Lambda}^\m{}_0 = \hat{p}^\m
        \,,\quad
        \tilde{\Lambda}^\m{}_a =  \Lambda^\m{}_a - \frac{(\hat{p}+\Lambda_0)^\m (\hat{p}\cdot \Lambda_a)}{  \hat{p} \cdot (\hat{p}+\Lambda_0)}
        \,,\\
        &\tilde{S}_{\m\n} = \hat{\eta}_{\m\rho} S^{\rho\sigma} \hat{\eta}_{\sigma\n} \,,\quad
        \tilde{x}^\m =  x^\m - S^{\m\n}\frac{p_\n}{|p|^2}
        \,.
    \end{split}
    \label{eq:spin-gauge-inv}
\end{align} 
The invariance of $\tilde{x}^\m$ and $\tilde{S}_{\m\n}$ can easily be checked against the flow equation. 
The change of $x^\m$ and $S_{\m\n}$ are such that 
$J_{\m\n} = x_\m p_\n - x_\n p_\m+ S_{\m\n}$ remains invariant. 

The invariance of $\tilde{\Lambda}$ has a nice geometric explanation \cite{Steinhoff:2015ksa}, 
which serves as a motivation behind the original spin-gauge generators $\mathcal{C}_\m$ in \eqref{eq:spin-gauge-stein}.
For any pair of time-like, unit-normalized, future-pointing 4-vectors $u^\m$ and $v^\m$, 
there exists a unique ``minimal boost" mapping one to the other: 
\begin{align}
    u^\m = B(u,v)^\m{}_\n v^\n\,,
    \quad 
    B(u,v)^\m {}_\n = \delta^\m{}_\n + \frac{(u+v)^\m (u+v)_\n}{1-u\cdot v} - 2u^\m v_\n \,. 
\end{align}
As emphasized in \cite{Steinhoff:2015ksa},
the relation between the two frames $\tilde{\Lambda}$ and $\Lambda$ in \eqref{eq:spin-gauge-inv} is nothing 
but the minimal boost between $\Lambda^\m{}_0$ and $\tilde{\Lambda}^\m{}_0 = \hat{p}^\m$:
\begin{align}
    \Lambda^\m{}_A = B(\Lambda_0,\tilde{\Lambda}_0)^\m{}_\n \tilde{\Lambda}^\n{}_A
    \quad 
    \Longleftrightarrow
    \quad 
    \tilde{\Lambda}^\m{}_A = B(\tilde{\Lambda}_0,\Lambda_0)^\m{}_\n \Lambda^\n{}_A \,.
    \label{eq:boost-L}
\end{align} 
Hence, if we solve the flow equation for $\Lambda^\m{}_0$, the solution for $\Lambda^\m{}_a$ will follow automatically 
from the minimal boost.

\paragraph{$\L$ orbit}

The equations for $\L$ do not depend on $S_{\m\n}$ or $x^\m$. 
So, we may solve them first and then feed the result into the other equations. 
The relation in \eqref{eq:spin-gauge-inv} implies 
\begin{align}
    \Lambda^\m{}_0 \tilde{\Lambda}_{\m a} = - \tilde{\Lambda}^\m{}_0 \Lambda_{\m a} = - \chi_a 
    \,\,\implies\,\,
    \Lambda^\m{}_0 = \sqrt{1+\vec{\chi}^2} \, \hat{p}^\m - \chi^a \tilde{\Lambda}^\m{}_a 
    \quad 
    (\vec{\chi}^2 = \chi^a \chi_a) \,.
\end{align}
Thus, the solution for $\chi^a(y)$ completely determines $\Lambda^\m{}_0$. The flow equation for $\chi^a$ 
away from the origin of the $y$-space is 
\begin{align}
    \partial_a \chi_b = 2 \{ \chi_b , \phi_a \} = \chi_a \chi_b + (1+ \sqrt{1+\vec{\chi}^2}) \delta_{ab} \,.
\end{align}
Noting that $\partial_a \chi_b - \partial_b \chi_a = 0$, 
we look for an SO$(3)$-covariant solution of the form 
\begin{align}
    \chi_a = \partial_a \left[ F(\vec{y}^2) \right] = 2y_a F'(\vec{y}^2) 
    \qquad 
    (\vec{y}^2 = y^a y_a) \,.
\end{align}
The equations admit a unique solution: $F' = (1-\vec{y}^2)^{-1}$.
In summary, the $\Lambda$ orbit is 
\begin{align}
    \Lambda^\m{}_B(y) = \tilde{\Lambda}^\m{}_A R^A{}_B(y) \,, 
    \quad 
    R^A{}_B(y) = \frac{1}{1-\vec{y}^2}
    \begin{pmatrix}
        1 + \vec{y}^2 & - 2 y_b
        \\
        - 2y^a & (1-\vec{y}^2) \delta^a{}_b + 2y^a y_b
    \end{pmatrix} \,.
    \label{L-sol}
\end{align}

\paragraph{$S$ and $x$ orbits} 
Intuitively, the invariant spin $\tilde{S}_{\m\n}$ accounts for the true rotation of the body, 
while the spin-gauge dependent part is induced by the boost \eqref{eq:boost-L}. In the gauge invariant frame, 
we split the components of the spin tensor as 
\begin{align}
    S_{\m\n}(y) = \tilde{\L}_\m{}^a \tilde{\L}_\n{}^b \tilde{S}_{ab} + 2 \tilde{\L}_{[\m}{}^0 \tilde{\L}_{\n]}{}^a U_a(y) \,.
    \label{S-split}
\end{align}
The flow equation for $U_a$ and its solution are simply 
\begin{align}
    \partial_a U_b = \tilde{S}_{ab} 
    \quad 
    \implies
    \quad
    U_a(y) = -  \tilde{S}_{ab} y^b \,.
    \label{S-sol-cov}
\end{align}
Finally, feeding this result into a relation in \eqref{eq:spin-gauge-inv}, we obtain the solution for $x^\m$: 
\begin{align}
    x^\m(y) = \tilde{x}^\m + S^{\m\n}(y) \frac{p^\n}{|p|^2} 
    = \tilde{x}^\m + \frac{1}{|p|}\tilde{\Lambda}^{\m a}\tilde{S}_{ab} y^b \,. 
\end{align}

\subsection{Physical phase space}

The goal of this subsection is to study the geometry of the 12-dimensional ``physical" or ``fully constrained" phase space. 
We use two slightly different but equivalent methods. 
First, we compute the Dirac brackets 
accounting for the four pairs of constraints. 
An advantage of the Dirac bracket is 
that we can keep using the original coordinates of the unconstrained phase space. 
Second, we choose new coordinates for the physical subspace 
and spell out the inclusion map into the unconstrained phase space. 
The spin-gauge invariant combinations \eqref{eq:spin-gauge-inv} 
strongly suggest an ideal choice of coordinates. 
Remarkably, we find that 
the physical phase space is a copy of $T^*(\mathrm{SO}(1,3))$ 
which is independent of the Regge trajectory. 

\paragraph{Dirac bracket} 
Appendix \ref{sec:constrained} 
gives a short review of Dirac brackets and related concepts. 
Applying the general methods to our model, 
we compute the ``fully constrained" Dirac brackets, 
\begin{align}
\{f , g \}_* 
:= \{ f , g\} -  \{ f , \phi_A \} \{ \chi^A , g \} + \{ f , \chi^A \} \{ \phi_A , g \}  \,.
\end{align} 
The complete list of non-vanishing Dirac brackets is as follows: 
\begin{align}
\begin{split}
    \{ x^\m , x^\n \}_* 
    &= \frac{1}{|p|^2} J^{\m\n} 
    \,,\\
    \{ x^\m , p_\n \}_* 
    &= \hat{\delta}^\m{}_\n 
    \,,\\
    \{ x^\m, \L_{\n}{}^a  \}_* 
    &=  \frac{1}{|p|} \left( 
        \hat{p}_\n \L^{\m a} - 2m' x^\m \tilde{S}_{\n\r} \Lambda^{\r a}
    \right)
    \,,\\
    \{ p_\m, \L_{\n}{}^a \}_* 
    &= 2m' \hat{p}_\m \tilde{S}_{\n\r} \Lambda^{\r a} 
    \,,\\
    \{ x^\r, S_{\m\n} \}_* 
    &=  -\frac{1}{|p|} \left( 
        \hat{p}_{\m} S_{\n}{}^\r - \hat{p}_{\n} S_{\m}{}^\r 
    \right)
    \,,\\
    \{ \Lambda_{\r}{}^a, S_{\m\n} \}_* 
    &=  -\left( 
        \hat{\eta}_{\r\m} \Lambda_{\n}{}^a - \hat{\eta}_{\r\n} \Lambda_{\m}{}^a 
    \right) 
    \,,\\ 
    \{ S_{\m\n} , S_{\r\s} \}_* 
    &= -\left( 
        \hat{\eta}_{\n\r} S_{\m \s} - \hat{\eta}_{\m\r} S_{\n\s} + \hat{\eta}_{\n\s} S_{\r\m} - \hat{\eta}_{\m\s}S_{\r\n} 
    \right) 
    \,. 
\end{split}
\label{Dirac-vector}
\end{align}
It is straightforward to confirm that the Dirac brackets satisfy the Leibniz rule and the Jacobi identity 
on the physical phase space. 
By construction, the Dirac brackets also satisfy 
\begin{align}
\{ \phi_A , f \}_* = 0 = \{ \chi^A , f \}_* \,,
\end{align} 
where $f$ is any function of the phase space coordinates. 

Finally, we note that
the difference between the Poisson bracket and the Dirac bracket may vanish
completely for some. 
For example, since all constraints are Lorentz invariant, 
\begin{align}
\{ J_{\m\n} , f \}_* = \{ J_{\m\n} , f \} \,. 
\end{align}
If both $f$ and $g$ are fully gauge invariant ($\{ f, \phi_A \} = 0 = \{ g , \phi_A \}$), 
then 
\begin{align}
    \{ f , g \}_* = \{ f , g \} \,. 
\end{align}

\paragraph{Minimal coordinates} 
The Dirac bracket \eqref{Dirac-vector} completely specifies 
the symplectic geometry of the physical phase space, 
but it may look opaque to an untrained eye. 
Here, we wish to introduce new coordinates on the physical phase space 
and spell out the inclusion map that solves the constraints explicitly. 

We demand that amongst the new coordinates 
the Dirac bracket and the Poisson bracket should coincide. 
We have seen some hints in this direction earlier. 
We recognize $\Lambda_\m{}^A$ and $J_{\m\n}$ as good candidates with 
$6+6$ components: 
\begin{align}
\label{eq:Dirac-LJ}
\begin{split}
    \{ \L_\m{}^A, \L_\n{}^B \}_* &= 0
    \,,
    \\
    \{ \L_{\r}{}^A, J_{\m\n}\}_* &= -( \eta_{\r\m} \L_{\n}{}^A - \eta_{\r\n} \L_{\m}{}^A )
    \,,
    \\ 
    \{ J_{\m\n}, J_{\r\s} \}_* &= -(\eta_{\n\r} J_{\m\s} - \eta_{\m\r} J_{\n\s} 
    + \eta_{\n\s} J_{\r\m} - \eta_{\m\s} J_{\r\n}) \,.
\end{split}
\end{align}
Since we are focusing on the physical phase space, 
$\Lambda$ here may be understood as the gauge  invariant $\tilde{\Lambda}$ 
discussed in section~\ref{sec:orbit}. 

The covariant spin-gauge condition enforces $\Lambda^\m{}_0 = \hat{p}^\m$. 
To see how this condition is implemented in the new coordinates, 
it is useful to define a pair of 4-vectors: 
\begin{align}
    \label{eq:VW-4vector}
    V_\m =  J_{\m\n}  \hat{p}^\n \,,
    \quad
    W_\m = - (*J)_{\m\n} \hat{p}^\n = - \frac{1}{2} \varepsilon_{\m\n\r\s} J^{\r\s} \hat{p}^\n  \,.
\end{align}
We recognize $W_\m$ as the normalized Pauli-Lubanski vector.
Substituting $J_{\m\n} = x_\m p_\n - x_\n p_\m + S_{\m\n}$ into \eqref{eq:VW-4vector} produces many terms,   
but most of them vanish on the physical phase space. 
In terms of the body frame coordinates  
$V_A =  V_\m \Lambda^\m{}_A$, $W_A =  W_\m \Lambda^\m{}_A$, 
we have
\begin{align}
    \label{eq:VW-body}
    \begin{split}
    V_{A=0} = 0 \,, &\quad V_a = - |p| x^\m \Lambda_{\m a} \,,
    \\
    W_{A=0} = 0  \,, &\quad W_a =  \frac{1}{2} \varepsilon_{a}{}^{bc} (S_{\m\n} \L^\m{}_b \L^\n{}_{c\,}) \,.
    \end{split}
\end{align}  
These expressions show how the three translation 
and three rotation degrees of freedom 
reside in the new coordinates. 
To clarify the physical meaning of $(\Lambda^\m{}_A, J_{\m\n})$ under constraints, 
we propose to use an equivalent but suggestive form $(\Lambda^\m{}_0, \Lambda^\m{}_a, V_a, W_a)$. 
The two sets are related by  $\Lambda^\m{}_0 = \hat{p}^\m$ and 
\begin{align}
    \begin{split}
    J_{a0} = L_{a0} = V_a = - m(W^2) x_a \,,
    \quad 
    J_{ab} = S_{ab} =  \varepsilon^{c}{}_{ab} W_c \,.
    \end{split}
\end{align}  
In other words, the new coordinates encode the physical degrees of freedom as
\begin{align}
    \begin{split}
    &\Lambda_\m{}^a \; :\; \mbox{(transverse rotation)} \,, \qquad J_{ab} = S_{ab} \; : \; \mbox{(spin angular momentum)} \,,
    \\
    &\Lambda_\m{}^0 \; :\; \mbox{(momentum direction)} \,, \quad \; J_{0a} = L_{0a} \; : \; \mbox{(transverse position)} \,.
    \end{split}
    \nonumber 
\end{align}
The inclusion map of the new coordinates $(\Lambda_\m{}^0, \Lambda_\m{}^a, V_a, W_a)$ 
into the original coordinates $(x^\m, p_\m, \L_\m{}^A, S_{\m\n})$ is simply given by
\begin{align}
    \label{eq:VW-embedding}
    x^\m = x_a \Lambda^{\m a} = - \frac{ V_a \Lambda^{\m a} }{m(W^2)}\,,
    \quad
    p_\m = m(W^2) \Lambda_{\m 0} \,,
    \quad 
    S_{\m\n} = \Lambda_\m{}^a \Lambda_\n{}^b \varepsilon^c{}_{ab} W_c \,.
\end{align}

\paragraph{Symplectic geometry} 
The above discussion can be rephrased in the language of symplectic geometry. 
The space-time components and body frame components are related 
in the usual way:
\begin{align}
    \begin{split}
        &p_A = p_\m \Lambda^\m{}_A \,,
        \quad 
        x^A = x^\m \Lambda_\m{}^A \,,
        \quad
        S_{AB} = S_{\m\n} \Lambda^\m{}_A \Lambda^\n{}_B \,,
        \quad 
        \\
        &\Theta^{AB} 
        = \Theta^{\m\n} \Lambda_\m{}^A \Lambda_\n{}^B 
        = (\Lambda^{\m C} d\Lambda^\n{}_C) \Lambda_\m{}^A \Lambda_\n{}^B 
        =  d\Lambda^{\n A} \Lambda_\n{}^B \,.
    \end{split}
\end{align}
The symplectic potential \eqref{symp-1} can be alternatively expressed as
\begin{align}
    \theta &= p_\m dx^\m + \frac{1}{2} S_{\m\n}\mem \Theta^{\m\n} = p_A dx^A + \frac{1}{2} J_{AB} \Theta^{AB} \,.
\end{align}
These two guises of $\theta$ naturally arise from a geometrical perspective; see \eqref{eq:MCforms-Poincare} and explanations thereafter.
Since the constraints break the SO$(1,3)_R$ 
 Lorentz symmetry acting on the body frame indices to SO$(3)_R$, 
we may write 
\begin{align}
    \theta &= p_0 dx^0 + p_a dx^a + (L_{0a} + S_{0a})\Theta^{0a} + \frac{1}{2}(L_{ab}+S_{ab})\Theta^{ab}\,.
\end{align}
The conditions $\chi^a=0$ set $p_a = 0$ 
and align the time axis of the body frame as $\Lambda_{\m 0} = \hat{p}_\m$.
The mass-shell constraint $\phi_0$ fixes the value of $p_{A=0}$ according to the Regge trajectory. 
The condition $\chi^0=0$ sets $x^{A=0}=0$. Finally, the conditions $\phi_a = 0$ set $S_{0a} = 0$. 

Imposing the constraints amounts to taking the pull-back of the symplectic potential 
onto the fully constrained phase space:
\begin{align}
    \theta_* &=  L_{0a} \Theta^{0a} + \frac{1}{2}S_{ab} \Theta^{ab}\,.
    \label{eq:theta-final}
\end{align}
Clearly, the second term carries the rotation degrees of freedom 
in the transverse 3-space.
As for the first term, we note that 
\begin{align}
    L_{0a} = -p_0 x_a = |p| x_a = m(\tilde{S}^2) x_a \,,
    \quad 
    \Theta^{0a} = d\Lambda^{\n 0} \Lambda_\n{}^a = - d(\hat{p}^\n) \Lambda_\n{}^a \,.
    \label{eq:boost-translation}
\end{align}
Mathematically, $L_{0a} \Theta^{0a}$ describes the boost part of the Lorentz transformation. 
Physically, it actually describes the translation degrees of freedom. 
To be concrete, $L_{0a}$ represents the transverse body frame components of the position $x^\m$, 
and $\Theta^{0a}$ measures the variation of the unit vector $\hat{p}$ in the body frame. 

In terms of $(\Lambda_\m{}^0, \Lambda_\m{}^a, V_a, W_a)$, 
the symplectic potential \eqref{eq:theta-final} on the physical phase space can also be written as 
\begin{align}
    \theta_* = - V_a \mem \Theta^{0a} + W_a  \left(\textstyle\frac{1}{2}\ve^a{}_{bc}\Theta^{bc}\right) \,.
\end{align}
The dependence on the Regge trajectory is nowhere to be seen.
This should come as no surprise 
since this symplectic potential carries precisely the same information 
as the Dirac bracket algebra \eqref{eq:Dirac-LJ} 
describing $T^*(\mathrm{SO}(1,3))$. 
Each point on the physical phase space labels a distinct trajectory (solution to the equations of motion).  
The ``shape" of the trajectory is not visible until time evolution, generated by $\phi_0$, begins to unfold. 
Only then will the Regge trajectory begin to play a crucial role.

\paragraph{Complex combinations}
For later purposes, we recombine $V_a$ and $W_a$ slightly. 
They implement $\mathfrak{so}(1,3)$ while keeping only the $\mathfrak{so}(3)$ covariance manifest:
\begin{align}
    \{ V_a , V_b \}_* = \ve^c{}_{ab} W_c
    \,,\quad
    \{ V_a , W_b \}_* = -\ve^c{}_{ab} V_c
    \,,\quad
    \{ W_a , W_b \}_* = -\ve^c{}_{ab} W_c
    \,.
\end{align}
Taking the standard complex combinations, 
\begin{align}
    J_a := \frac{1}{2} (W_a - iV_a)\,, 
    \quad
    \bar{J}_a := \frac{1}{2} (W_a + iV_a) \,, 
\end{align}
we may rewrite \eqref{eq:Dirac-LJ} as
\begin{align}
    \label{eq:VW-DB}
    \begin{split}
        &\{ J_a , J_b \}_* = -\ve^c{}_{ab} J_c \,,
        \\
        &\{ \Lambda_\m{}^0 , J_c \}_* = - \frac{1}{2i} \Lambda_{\m c} \,,
        \quad
        \{ \Lambda_\m{}^a , J_c \}_* = - \frac{1}{2i} \delta^a{}_c \Lambda_\m{}^0 -\frac{1}{2}  \varepsilon^a{}_{cb} \Lambda_\m{}^b \,,
    \end{split}
\end{align}
and their complex conjugates. 
Combining the Dirac bracket \eqref{eq:VW-DB} and the inclusion map \eqref{eq:VW-embedding} 
offers perhaps the most intuitive way to understand the Dirac bracket 
computed directly in the original coordinates \eqref{Dirac-vector}.

\paragraph{From vector to spinor} 
Anticipating the comparison with the massive twistor model in the next section, 
we convert the vector indices into spinor indices. 
The Lorentz indices $(\m,\n,\ldots)$ are converted to SL$(2,\mathbb{C})$ bi-spinor indices 
$(\a,\b,\ldots ; \da,\db, \ldots)$. 
A body frame vector $X_A$ is converted to a little group $\mathfrak{su}(2)$ bi-spinor $X_{IJ}$ such that 
the anti-symmetric part $X_{[IJ]}$ carries the $X_{A=0}$ component and the symmetric part $X_{(IJ)}$
carries the $X_{a=1,2,3}$ components. 
See appendix \ref{sec:spinor-conv} for a detailed summary of our conventions for spinors and twistors, 
and the precise conversion rules between vector indices and spinor indices. 

We may apply the conversion rules either to the body frame coordinates $(V_a, W_a, \Lambda_\m{}^A)$ parametrizing 
the constrained subspace 
or to the original coordinates $(x^\m, p_\m , \Lambda_\m{}^A, S_{\m\n})$ parametrizing 
the unconstrained phase space. We convert the common part first, 
\begin{align}
    \Lambda^{IJ}_{\a\da} = (\sigma^\m)_{\a\da}  \Lambda_\m{}^A (\sigma_A)^I{}_J \,.
    \label{eq:Lambda-spinor}
\end{align}
The conversion rule for the body frame vectors $V$ and $W$ is (see \eqref{eq:SU2-A-to-IJ})
\begin{align}
    X_{IJ} = \s^A_{IJ} X_A \,.
\end{align}
In the spinor notation, the Dirac bracket \eqref{eq:VW-DB} can be rewritten as
\begin{align}
\label{eq:Dirac-LJ-SU2}
\begin{split}
    \{ J^I{}_J, J^K{}_L \}_* &= -i ( \delta^I{}_L J^K{}_J - \delta^K{}_J J^I{}_L )
    \,,\\
    \{ (\Lambda_{\a\da})^I{}_J , J_{KL} \}_* &= + i (\L_{\a\da})^I{}_{(K}\epsilon_{L)J} 
    \,.
\end{split}
\end{align}
The conversion rules for $(x,p,S)$ are as follows: 
\begin{align}
    p_{\a\da} = (\sigma^\m)_{\a\da} p_\m \,,
    \quad
    x^{\da\a} = (\bar{\sigma}_\m)^{\da\a} \left(\textstyle{\frac{1}{2}} x^\m \right) \,,
    \quad 
    \textstyle{\frac{1}{2}} S_{\m\n} \sigma^{[\m}_{\a\da} \sigma^{\n]}_{\b\db} 
    = S_{\a\b} \bar{\epsilon}_{\da\db} + \epsilon_{\a\b} \bar{S}_{\da\db} \,. 
\end{align}
The reality conditions are recast as 
\begin{align}
    [p_{\a\db}]^* = p_{\b\da} \,,
    \quad
    [x^{\da\b}]^* = x^{\db\a} \,,
    \quad 
    [ \Lambda^{IJ}_{\alpha\dot{\beta}} ]^* =  - \Lambda_{JI,\beta\dot{\alpha}} \,,
    \quad
    [S_{\a\b}]^* = \bar{S}_{\da\db} \,.
\end{align}
The non-vanishing Dirac brackets in \eqref{Dirac-vector} are rewritten as
\begin{align}
    \begin{split}
    \{ x_{\a\da}, x_{\b\db} \}_* &= \frac{1}{2|p|^2} \left( J_{\a\b} \be_{\da\db} + \e_{\a\b} \bar{J}_{\da\db} \right)
    \,,\\
    \{ x_{\a\da} , p_{\b\db} \}_* &= - \left( 
        \e_{\a\b} \be_{\da\db} - \hat{p}_{\alpha\dot{\alpha}} \hat{p}_{\beta\dot{\beta}} 
    \right)
    \,,\\
    \{ x_{\a\da} , \Lambda^{IJ}_{\b\db}   \}_* &= \frac{1}{2|p|} \left(
        \hat{p}_{\b\db} \L^{IJ}_{\a\da} 
        + 4m' x_{\a\da} \big(
            S_{\b\c}\e^{\c\d}\L_{\d\db}^{IJ}
            + \bar{S}_{\db\dc} \be^{\dc\dd}\L_{\b\dd}^{IJ}
        \big)
    \right)
    \,,\\
    \{ p_{\a\da} , \Lambda^{IJ}_{\b\db}  \}_* &= -2m'\hat{p}_{\a\da} \big(
        S_{\b\c}\e^{\c\d}\L_{\d\db}^{IJ}
        + \bar{S}_{\db\dc}\be^{\dc\dd}\L_{\b\dd}^{IJ}
    \big)
    \,,\\
    \{ x_{\g\dg}, S_{\a\b} \}_* 
    &= \frac{1}{4|p|}\big( \hat{p}_{\a\dg} S_{\b\g} + \hat{p}_{\b\dg} S_{\a\g} - \hat{p}_{\g\dg} S_{\a\b} \big)
    \,,\\
    \{ \Lambda^{IJ}_{\g\dg}, S_{\a\b} \}_* 
    &= \frac{1}{2} \left( 
        \e_{\a\g} \Lambda^{IJ}_{\b\dg} + \e_{\b\g}  \Lambda^{IJ}_{\a\dg} 
        - \frac{1}{2} \hat{p}_{\g\dg} \big(
            \L^{IJ}_{\a\da} \hat{p}_{\b\db} + \L^{IJ}_{\b\da} \hat{p}_{\a\db} 
        \big) \be^{\da\db}
    \right) 
    \,,\\
    \{ S_{\a\b}, S_{\g\d} \}_* &= - \frac{1}{2} \big( \e_{\a\g} S_{\b\d} + \e_{\b\g} S_{\a\d} + \e_{\a\d} S_{\b\g} + \e_{\b\d} S_{\a\g} \big)
    \,,\\
    \{ S_{\a\b}, \bar{S}_{\dg\dd} \}_* &= 
    - \frac{1}{2} \left( \hat{p}_{\a\dg} (S\cdot \hat{p}) _{\b\dd} + \hat{p}_{\b\dg} (S\cdot \hat{p})_{\a\dd} 
    + \hat{p}_{\a\dd} (S\cdot \hat{p}) _{\b\dg} + \hat{p}_{\b\dd} (S\cdot \hat{p})_{\a\dg} \right) 
    \,,
\end{split}
\label{Dirac-spinor}
\end{align} 
while the covariant gauge condition \eqref{cov-gauge-old} is rephrased as
\begin{align}
    S_{\m\n} p^\n = 0 
    \quad 
    \Longleftrightarrow
    \quad
    S_{\a\b} \epsilon^{\b\g} p_{\g\da} = p_{\a\db} \bar{\epsilon}^{\db\dg} \bar{S}_{\dg\da} \,,
\end{align}
i.e., the bi-spinor $(S\cdot p)_{\a\da} := S_{\a\b} \epsilon^{\b\g} p_{\g\da}$ as a $(2\times 2)$ matrix should be anti-hermitian.

\subsection{Equation of motion} \label{sec:vect-eom}

We study the equations of motion of the spherical top model. 
A covariant Hamiltonian description of a relativistic particle 
is somewhat subtle due to the fact the time evolution generator $\phi_0$ 
is a gauge generator at the same time. 
We circumvent the subtlety by treating $(\phi_0,\chi^0)$ differently from $(\phi_a, \chi^a)$, 
although it is possible to put all gauge generators on an equal footing; see {\it e.g.} ref.~\cite{Barandes:2019oas}.

\paragraph{Without spin} 

Let us quickly review the equation of motion of a spin-less particle from a Hamiltonian point of view. 
The first-order action is 
\begin{align}
\mathcal{S} = \int d\sigma \left( p_\mu \dot{x}^\m - \frac{\kappa}{2}(p^2 + m^2) \right) \,,
\end{align} 
where $\sigma$ is a world-line parameter and $\dot{x} = dx/d\sigma$. The mass-shell constraint is imposed by the Lagrange multiplier $\kappa$. 
The variations of the action give
\begin{align}
\begin{split}
\delta_x \mathcal{S} = 0 
\quad 
&\Longrightarrow
\quad 
\dot{p}_\mu = 0 \,,
\\
\delta_p \mathcal{S} = 0 
\quad 
&\Longrightarrow
\quad 
\dot{x}^\mu - \kappa p^\m = 0 \,, 
\\
\delta_\kappa \mathcal{S} = 0 
\quad 
&\Longrightarrow
\quad 
p^2 + m^2 = 0 \,.
\end{split}
\end{align} 
Combining the equations, we can determine $\kappa$:
\begin{align}
\kappa = \frac{1}{m} \frac{d\tau}{d\sigma} \,,
\quad d\tau = \sqrt{-dx^\mu dx_\mu} \,.
\label{multi-fixed}
\end{align}
Note that the constraint $\phi_0$ in the action plays the role of the Hamiltonian generating the time evolution, 
even though its numerical value stays zero. 

\paragraph{With spin}
Since $\phi_a$ form first-class constraints, we can use either of the two equivalent ways to derive the equation of motion under the constraints. One is to add a linear combination of $\phi_a$ in the action using Lagrange multipliers. The other is to use the Dirac bracket constructed from the pair $(\phi_a,\chi^a)$. For illustrative purposes, we examine both approaches in some detail.

For the first approach, we begin with the action:
\begin{align}
\begin{split}
\mathcal{S} &= \int d\sigma \left( p_\mu \dot{x}^\m + \frac{1}{2} S_{\m\n} \Omega^{\m\n} - \kappa \phi_0 - \kappa^a \phi_a \right)
\\
&= \int d\sigma \left( p_\mu \dot{x}^\m + \frac{1}{2} S_{\m\n} \Omega^{\m\n} 
- \frac{\kappa}{2}(p^2 + m^2) 
- \frac{\kappa^a}{2} (\hat{p}^\mu + \Lambda^\mu{}_0) S_{\mu\nu} \Lambda^\n{}_a \right) \,.
\end{split}
\end{align} 
The angular velocity is defined as the pull back of the one-form $\Theta^{\m\n}$ onto the world-line: 
\begin{align}
\label{eq:angvel-def}
\Omega^{\m\n} =  \L^{\m A} \frac{d \L^\n{}_A}{d\s} \,.
\end{align}
The variations of the Lagrange multipliers $\kappa^0$, $\kappa^a$ force the gauge generators $\phi_0$, $\phi_a$ to vanish. 
The variations of the dynamical variables yield the equations of motion:
\begin{align}
\begin{split}
\delta_x \mathcal{S} = 0 
\quad 
&\Longrightarrow
\quad 
\dot{p}_\mu = 0 \,,
\\
\delta_p \mathcal{S} = 0 
\quad 
&\Longrightarrow
\quad 
\dot{x}_\mu - \kappa \left(p_\m - \frac{2m m'}{|p|^2} \tilde{S}_{\m\n}  S^{\n\r} p_\r \right) 
- \frac{\kappa^a}{2|p|}  \hat{\eta}_{\mu\nu} S^\nu{}_\rho \Lambda^\rho{}_a  = 0 \,, 
\\
\delta_S \mathcal{S} = 0 
\quad 
&\Longrightarrow
\quad 
\Omega^{\m\n} - 2 \kappa m m' \tilde{S}^{\m\n} - \kappa^a  (\hat{p} + \Lambda_0)^{[\m} \Lambda^{\n]}{}_a = 0 \,.
\end{split}
\label{eom-vector}
\end{align}
The variation of $\Lambda_A{}^\mu$ takes values in $\mathfrak{so}(1,3)$:
\begin{align}
\delta \L^\m{}_A = -\delta{\theta}^{\m}{}_\n \L^{\n}{}_A   \,, 
\quad 
\delta \theta^{\n\m} = - \delta{\theta}^{\m\n} \,.
\end{align}
The intermediate steps include
\begin{align}
\begin{split}
\delta_{\L} \left( S_{\m\n} \O^{\m\n} \right) 
&= -\delta \theta^{\m\n} \left(   \dot{S}_{\m\n}  - 2 S_{\rho[\m} \Omega_{\n]}{}^\rho \right) \,,
\\
\delta_{\L} \left[ (\hat{p}+ \Lambda_0)^\m S_{\mu\nu}\Lambda^\nu{}_a  \right]
&= -\delta \theta^{\m\n} \left(   (\hat{p}+\L_0)^\rho S_{\rho[\m} \L_{\n] a} - \L_{0[\m} S_{\n]\rho} \L^\rho{}_a  \right)\,.
\end{split}
\end{align}
Combining all terms, we find an equation for $\dot{S}$: 
\begin{align}
\dot{S}_{\m\n} = 2 S_{\rho [\m} \Omega_{\n]}{}^\rho 
+ \kappa^a  (\hat{p}+\L_0)^\rho S_{\rho[\m} \L_{\n] a} -  \kappa^a \L_{0[\m} S_{\n]\rho} \L^\rho{}_a  \,. 
\label{eom-Sdot}
\end{align}
Before taking the next step, let us collect all the equations of motion in one place:
\begin{align}
\begin{split}
\dot{p}_\mu &= 0 \,,
\\
\dot{x}_\mu &= \kappa \left( p_\m - \frac{2m m'}{|p|^2} \tilde{S}_{\m\n}  S^{\n\r} p_\r \right) + \frac{\kappa^a}{2|p|}  \hat{\eta}_{\mu\nu} S^\nu{}_\rho \Lambda^\rho{}_a  \,, 
\\
\Omega^{\m\n} &= 2 \kappa m m' \tilde{S}^{\m\n} + \kappa^a  (\hat{p} + \Lambda_0)^{[\m} \Lambda^{\n]}{}_a \,, 
\\
\dot{S}_{\m\n} &= 2 S_{\rho [\m} \Omega_{\n]}{}^\rho + \kappa^a  (\hat{p}+\L_0)^\rho S_{\rho[\m} \L_{\n] a} -  \kappa^a \L_{0[\m} S_{\n]\rho} \L^\rho{}_a  \,. 
\end{split}
\label{eom-all}
\end{align}

\paragraph{Lagrange multipliers} 

As shown in \eqref{multi-fixed}, 
The Lagrange multiplier $\kappa$ for the mass-shell constraint is related to the world-line reparametrization invariance. The multipliers $\kappa^a$ for the spin-gauge symmetry are uniquely fixed by the gauge-fixing conditions; see appendix~\ref{sec:constrained} for the relation between the multipliers, gauge-fixing conditions and Dirac bracket.

Our favorite gauge-fixing condition is the covariant one introduced in \eqref{def:cov-gauge-fixing}. 
Requiring that the gauge-fixing condition is preserved by the equations of motion, we find
\begin{align}
\begin{split}
0 &= \frac{d}{d\sigma} \left( \hat{p}_\nu \Lambda^\nu{}_a \right)
\\
&= -   \hat{p}_\mu \Omega^{\mu\nu} \L_{\nu a}
\\
&=  - \hat{p}_\mu \left( 2 \kappa m m' \tilde{S}^{\m\n} + \kappa^b (\hat{p} + \Lambda_0)^{[\m} \Lambda^{\nu]}{}_b \right)  \L_{\nu a}
\\
&\approx - \frac{1}{2}  \kappa^b \hat{p}_\mu \left(  (\hat{p} + \Lambda_0)^{\m} \Lambda^{\n}{}_b - (\hat{p} + \Lambda_0)^{\n} \Lambda^{\m}{}_b \right)  \L_{\nu a}
= \eta_{ab} \kappa^b \,.
\end{split}
\label{mult-free}
\end{align}
In the last line, we took the liberty to impose both $\phi_A = 0$ and $\chi^A=0$. 

We reach the expected conclusion that, for a free particle, the multipliers $\kappa^a$ all vanish. 
The final form of the equations of motion describes a motion with constant (proper)
linear velocity and angular velocity. 
\begin{align}
\begin{split}
&\frac{dp_\mu}{d\tau} = 0 \,,
\quad 
p^\mu = m \frac{dx^\mu}{d\tau}  \,, 
\\
&\frac{d S_{\m\n}}{d\tau} = 2 S_{\rho [\m} \bar{\Omega}_{\n]}{}^\rho = 0 \,,
\quad
S^{\m\n} = \frac{1}{2 m'} \bar{\Omega}^{\m\n} \,,
\quad 
\left( \bar{\Omega}^{\m\n} := \L^{\m A} \frac{d\L^\n{}_A}{d\tau} \right)\,, 
\\
&S_{\m\n} p^\n = 0 \,, 
\quad 
\L^\n{}_0 = \hat{p}^\n \,.
\end{split}
\label{eq:eom-all-tau}
\end{align}
In the Newtonian limit, the Regge trajectory can be approximated by
\begin{align}
    \begin{split}
    &m(\tilde{S}^2) = m_0 + \frac{1}{2I} \tilde{S}^2 + \cdots 
    \quad 
    \left( m'(\tilde{S}^2)=\frac{1}{2I} + \cdots \right)
    \\
    \implies
    \quad 
    &p^0 
    = \sqrt{\vec{p}^2 +m^2}
    = m_0 + \frac{1}{2m_0} {\vec{p}\mem}^2 + \frac{1}{2I} \tilde{S}^2 + \cdots\,,    
    \end{split}
\end{align}
such that $S^{\m\n} = (1/2m') \bar{\Omega}^{\m\n}$ in \eqref{eq:eom-all-tau} reduces to 
the familiar $\vec{S} = I \,\vec{\Omega}$.

\paragraph{Dirac bracket revisited}

When first-class constraints $\phi_a$ are paired with conjugate constraints $\chi^a$ to form a set of second-class constraints, 
Dirac brackets can be used to derive the equations of motion without writing out the full action with Lagrange multipliers. 
To apply the method to our model, we construct the ``partly constrained" Dirac brackets, 
\begin{align}
\{f , g \}_\diamond := \{ f , g\} -  \{ f , \phi_a \} \{ \chi^a , g \} + \{ f , \chi^a \} \{ \phi_a , g \} 
 \,.
\end{align} 
Here, we say ``partly constrained" to emphasize the fact that we left out the mass-shell constraints $(\chi^0, \phi_0)$.
The complete list of non-vanishing Dirac brackets among phase space coordinates is as follows: 
\begin{align}
\begin{split}
\{ x^\mu , x^\nu \}_\diamond &= \frac{1}{|p|^2} S^{\m\n} \,,
\\
\{ x^\mu , p_\nu \}_\diamond &= \delta^\m{}_\n \,,
\\
\{ x^\mu, \L_{\nu}{}^a \}_\diamond &= \frac{1}{|p|^2} p_\nu \L^{\m a} \,,
\\
\{ x^\rho , S_{\mu\nu} \}_\diamond &=  -\frac{1}{|p|^2} \left( p_\mu S_\nu{}^\rho - p_\nu S_\mu{}^\rho \right)\,,
\\ 
\{ \Lambda_{\rho}{}^a , S_{\mu\nu}\}_\diamond &= - ( \hat{\eta}_{\r\m} \Lambda_{\nu}{}^a - \hat{\eta}_{\r\n} \Lambda_{\mu}{}^a )\,,
\\ 
\{ S_{\mu\nu} , S_{\rho\sigma} \}_\diamond &= - (\hat{\eta}_{\nu\rho} S_{\mu \sigma} - \hat{\eta}_{\mu\rho} S_{\nu\sigma} + \hat{\eta}_{\nu\sigma} S_{\rho\mu} - \hat{\eta}_{\mu\sigma}S_{\rho\nu}) \,. 
\end{split}
\label{Dirac-spatial-only}
\end{align}
We did not include the time-like frame vector ($\Lambda^\m{}_0$) in the list, 
because the covariant gauge condition enforces $\Lambda^\m{}_0 = \hat{p}^\m$, 
which in turn implies $\{ \Lambda^\m{}_0 , X \}_\diamond = \{ \hat{p}^\m, X \}_\diamond$ 
for any phase space coordinate $X$. 

The Dirac bracket guarantees that both the gauge generators and the gauge-fixing conditions are preserved throughout the Hamiltonian time evolution.
\begin{align}
\begin{split}
\{ \phi_a , h \}_\diamond 
&= \{ \phi_a , h \} -  \{ \phi_a , \phi_b \} \{ \chi^b , h \} + \{ \phi_a , \chi^b \} \{ \phi_b , h \}  
\\
&=\{ \phi_a , h \} -  0 -\delta_a{}^b \{ \phi_b , h \} = 0 \,,  
\\
\{ \chi^a , h \}_\diamond 
&= \{ \chi^a , h \} -  \{ \chi^a , \phi_b \} \{ \chi^b , h \} + \{ \chi^a , \chi^b \} \{ \phi_b , h \} 
\\
&= \{ \chi^a , h \} -  \delta^a{}_b \{ \chi^b , h \} + 0 = 0 \,. 
\end{split}
\end{align}

Using $\phi_0$ as an effective Hamiltonian for the time evolution, we arrive at a general prescription for the equation of motion: 
\begin{align}
\frac{df}{d\sigma} = \kappa^0 \{ f, \phi_0 \}_\diamond 
\qquad 
\mbox{or}
\qquad
\frac{df}{d\tau} = \frac{1}{m} \{ f, \phi_0 \}_\diamond \,.
\label{eom-dirac}
\end{align}
Inserting $x$, $p$ and $S$, we immediately recover
\begin{align}
    \frac{dx^\m}{d\tau} = \frac{1}{m} p^\m \,,
    \quad 
    \frac{dp^\m}{d\tau} = 0 \,,
    \quad 
    \frac{d S_{\m\n}}{d\tau} = 0 \,.
\end{align} 
Inserting $\Lambda$, we find
\begin{align}
    \frac{d\Lambda^\n{}_a}{d\tau} =  2 m' \tilde{S}^{\r\s} \delta_{[\s}^\n \Lambda_{\r] a} 
    \quad
    \implies
    \quad 
    \bar{\Omega}^{\m\n} = 2m'\mem S^{\m\n} \,.
\end{align} 
Thus, the Dirac bracket method reproduces all equations of motion in \eqref{eq:eom-all-tau} as expected.

\section{Massive twistor} \label{sec:spin}

We introduce the massive twistor model \cite{Fedoruk:2014vqa} in such a way that 
its equivalence to the spherical top model appears almost inevitable. 
In the previous section, 
we showed that the physical space of the spherical top model is a copy of $T^*(\mathrm{SO}(1,3))$ 
and that the Regge trajectory carries all the dynamical information not dictated by the symmetries. 

Our goal in this section is clear from the beginning. 
First, we should show that the physical phase space of the twistor model 
is also a copy of $T^*(\mathrm{SO}(1,3))$. 
Second, we should show how to incorporate the Regge trajectory into the 
twistor model.
Finally, the symmetries and the Regge trajectory should be compatible 
such that the solutions to the equations of motion 
agree perfectly between the two models.

\subsection{Massive twistor with constraints}

We assume that the readers are somewhat familiar with the massless twistor space. 
Our conventions for the twistor space and its symmetry algebra are summarized 
in the second half of section~\ref{sec:spinor-conv}. 
We follow the same conventions whenever applicable and focus on the differences between the massive case and the massless case.

\paragraph{Twistor space and its symmetry} 

The massive twistor consists of a doublet of massless twistors: $Z_\mathrm{A}{}^I$ ($\mathrm{A}=1,2,3,4$, $I=1,2$). 
The Lie algebra of the natural symmetry group acting on this space is 
$\mathfrak{su}(2,2)\oplus \mathfrak{su}(2) \oplus \mathfrak{u}(1)$. 
The mass of the particle partially breaks the conformal algebra 
$\mathfrak{su}(2,2)$
and leave behind the Poincar\'e subalgebra.
The ``internal''  $\mathfrak{su}(2)$ is to be identified with the little group symmetry 
of the spherical top model.
We will not discuss the possibility of associating the $\mathfrak{u}(1)$ with an electric charge. 

The twistor coordinates and their conjugations with respect to the SU$(2,2)$ invariant tensor are denoted by 
\begin{align}
    Z_{\mathrm{A}}{}^I = 
    \begin{pmatrix}
        \lambda_\a{}^I
        \\
        i \mu^{\da I}
    \end{pmatrix}
    \,,\quad
    \bar{Z}_I{}^{\mathrm{B}} := (Z_{\mathrm{A}}{}^I)^\dagger A^{\bar{\mathrm{A}}\mathrm{B}} = 
    \begin{pmatrix}
        -i \bar{\mu}_I{}^{\b}
        &
        \bar{\lambda}_{I\db}
    \end{pmatrix}
    \,.
\end{align}
The symplectic form invariant under the $\mathfrak{su}(2,2)\oplus \mathfrak{su}(2) \oplus \mathfrak{u}(1)$ is given by 
\begin{align}
    \omega 
    = i (d\bar{Z}_I{}^{\mathrm{A}} \wedge d Z_{\mathrm{A}}{}^{I}) 
    = d \bar{\mu}_I{}^{\alpha} \wedge d\lambda_{\alpha}{}^I - d\bar{\lambda}_{I\dot{\alpha}}\wedge d\mu^{\dot{\alpha} I} \,.
\end{align}
The Poisson brackets are 
\begin{align}
    \{ Z_{\mathrm{A}}{}^I, \bar{Z}_J{}^{\mathrm{B}} \} = - i\, \delta_\mathrm{A}{}^\mathrm{B} \delta^I{}_J 
    \,\,\implies\,\,
    \{\lambda_{\alpha}{}^I , \bar{\mu}_J{}^{\beta} \} =  \delta_{\alpha}{}^{\beta} \delta^I{}_J
        \,,\quad
    \{\bar{\lambda}_{I \dot{\alpha}}, \mu^{\dot{\beta}J}\} =\delta_{\dot{\alpha}}{}^{\dot{\beta}} \delta_I{}^J \,.
\end{align}
The generators of Hamiltonian flows for $\mathfrak{su}(2,2)$ are 
\begin{align}
 G_{\mathrm{A}}{}^{\mathrm{B}} := 
 Z_{\mathrm{A}}{}^I \bar{Z}_I{}^{\mathrm{B}}  - \frac{1}{4}  \delta_\mathrm{A}{}^\mathrm{B} (\bar{Z}_I{}^{\mathrm{C}} Z_{\mathrm{C}}{}^I) \,.
    \label{eq:su22-massive}
\end{align}
They are identified with the conformal generators as
\begin{align}
    \begin{split}
        G_\mathrm{A}{}^\mathrm{B}  =  
        \begin{pmatrix}
            -i J_\a{}^\b - \textstyle{\frac{i}{2}} \delta_\a{}^\b D  & - P_{\a\db} 
            \\ 
            - K^{\da\b} &  +i \bar{J}^\da{}_{\db}  + \textstyle{\frac{i}{2}} \delta^\da{}_{\db} D  
        \end{pmatrix} \,,
    \end{split}
    \label{eq:su22-conf-massive}
\end{align}
such that 
\begin{align}
    \begin{split}
        J_{\a\b} = \bar{\mu}_{I(\a} \lambda_{\b)}{}^I \,, 
        \quad
        &D =  \frac{1}{2} \left( \langle\lmu\lambda\rangle + [\rambda\rmu] \right)  \,,
        \\
        P_{\a\da} = - \lambda_\a{}^I \bar{\lambda}_{I \da} \,,
        \quad 
        &K^{\da\a} = - \mu^{\da I} \bar{\mu}_I{}^\a \,,
    \end{split}
    \label{eq:conf-twistor-massive}
\end{align}
where we defined
\begin{align}
    \langle\lmu\lambda\rangle 
    := \lmu_I{}^\a \lambda_\a{}^I
    \,,\quad
    [\rambda\rmu] 
    := \rambda_{I\da} \rmu^{\da I}
    \,.
\end{align}
The generators of Hamiltonian flows for $\mathfrak{su}(2)\oplus\mathfrak{u}(1)$ are 
\begin{align}
    W^I{}_J 
    = Z_{\mathrm{A}}{}^I \bar{Z}_J{}^{\mathrm{A}}
    \,\,\implies\,\,
    \{ W^I{}_J, W^K{}_L \} = 
    {-} i ( \delta^I{}_L W^K{}_J - \delta^K{}_J W^I{}_L )
    \,.
    \label{su2-twistor}
\end{align}

\paragraph{Constraints on the twistor space}

Before imposing constraints, the massive twistor space carries 16 real (8 complex) degrees of freedom. 
To establish the equivalence between the two models, 
at the level of the fully constrained physical phase space with 12 real physical degrees of freedom,
we should introduce 2 real (1 complex) gauge generators and the corresponding gauge-fixing conditions.

The twistor model has no analogue of the spin-gauge symmetry of the spherical top model. 
In a rough sense, the twistor model offers a solution to the covariant spin condition.
So, we only need to consider the twistor incarnation of the mass-shell constraint. 
Recall the mass-shell constraint and its conjugate from the spherical top model:
\begin{align}
    \phi_0 = \frac{1}{2}\left( p^2 + m^2(\tilde{S}^2) \right) \,, \quad 
    \chi^0 = \frac{1}{p^2} x^\mu p_\mu  \,.
    \nonumber 
\end{align}
We propose analogous pairs of constraints in the twistor model: 
\begin{align}
\begin{split}
    \label{spinor-constraint}
    \phi &= -\frac{1}{2} \left( \det(\lambda) - m(\tilde{S}^2) \right) 
    \,,\quad 
    \bar{\chi} 
    = \frac{1}{\det(\lambda)} \langle\lmu\lambda\rangle 
    \,,
    \\
    \bar{\phi} &= -\frac{1}{2} \left( \det(\bar{\lambda}) - m(\tilde{S}^2) \right) 
    \,,\quad 
    \chi 
    = \frac{1}{\det(\rambda)} [\rambda\rmu]    
    \,.
\end{split}
\end{align}
This proposal is natural in many ways. 
First, the condition $\det(\lambda) = m$ takes a ``square-root'' of the mass shell condition $-p^2 = m^2$.  
A slight difference is that the twistor version fixes not only the magnitude but also the phase of $\det(\lambda)$.
Second, just as $\phi_0$ in the spherical top model generates the ``time gauge orbit' 
in the real Minkowski space, $x^\m \mapsto x^\m + k p^\m$, 
$(\phi,\bar{\phi})$ in the twistor model generates the time gauge orbit 
in the complexified Minkowski space, $z^\m \mapsto z^\m + \kappa p^\m$.
Third, while $\chi^0$ in the spherical top model projects the $x$-space 
onto the real three-dimensional  subspace transverse to $p^\m$, 
$(\bar{\chi},\chi)$ in the twistor model projects the $z$-space 
onto the complex three-dimensional transverse subspace. 
Finally, the constraints have been unit-normalized: 
\begin{equation}
    \{\bar{\chi}, \phi\} = 1
    \,,\quad
    \{\chi, \bar{\phi}\} = 1
    \,.
\end{equation}
Readers unfamiliar with the use of complexified Minkowski space in the twistor model 
might have been puzzled by some of the statements above. 
More information will be given.

\paragraph{Map to the spherical top model} 
Strictly speaking, the map between the two models is 2-to-1 
for the same reason as $\mathrm{SO}(3) = \mathrm{SU}(2)/\mathbb{Z}_2$. 
We will ignore this double cover issue in most of what follows.

The map from the twistor model to the spherical top model 
can be phrased either in terms of the original coordinates $(x^\m, p_\m, \Lambda^\m{}_A, S_{\mu\nu})$ 
or in terms of the physical coordinates $(\Lambda^\m{}_0,\Lambda^\m{}_a,V_a, W_a)$. 
We find it instructive to spell out both descriptions, 
beginning with the common factor: $\Lambda^\m{}_A \in \mathrm{SO}(1,3)$.
The twistor counterpart of $\Lambda^\m{}_A$ involves only $(\lambda,\bar{\lambda})$ and 
not $(\mu, \bar{\mu})$. This is natural in that $\Lambda$ and $(\lambda,\bar{\lambda})$ 
span Lagrangian submanifolds of their respective phase spaces 
both before and after imposing constraints.

\paragraph{Momentum and body frame}
The map for $p_\mu$ is waiting for us in \eqref{eq:conf-twistor-massive}:
\begin{align}
    p_{\alpha\dot{\alpha}} = - \lambda_{\alpha}{}^I \bar{\lambda}_{I \dot{\alpha} }
    \,. 
\label{map-p}
\end{align} 
The overall minus sign is due to the fact that $p_0 <0$ in our convention. 
The reality of $p_\mu$ is consistent with the fact that $\lambda$ and $\bar{\lambda}$ are hermitian conjugates. 

In the covariant gauge, $\Lambda^\m{}_A$ is aligned such that $\Lambda^\mu{}_0 = \hat{p}^\mu$, 
while the magnitudes of the momentum enter the Regge trajectory in a Lorentz invariant way. 
In an analogous way, we can separate the magnitudes of the twistor variables by 
\begin{align}
   \det(\lambda)\mem \e^{IJ} := \e^{\a\b} \lambda_\a{}^I \lambda_\b{}^J   \,,
   \quad 
    \det(\bar{\lambda}) := [\det(\lambda)]^* \,,
\end{align}
and define the normalized variables as
\begin{align}
\begin{split}
    \label{eq:twistor-normalized}
    \hat{\lambda}_\a{}^I := (\det(\lambda))^{-1/2} \lambda_\a{}^I
    &\,,\quad
    \hat{\rmu}^{\da I} := (\det(\lambda))^{1/2} \rmu^{\da I}
    \,,\\
    \hat{\rambda}_{I\da} := (\det(\rambda))^{-1/2} \rambda_{I\da}
    &\,,\quad
    \hat{\lmu}_I{}^\a := (\det(\rambda))^{1/2} \lmu_I{}^\a
    \,.\\
\end{split}
\end{align}
The map from $(\hat{\lambda},\hat{\bar{\lambda}})$ to $\Lambda$, which includes the normalized version of \eqref{map-p}, 
is self-evident, because $(\hat{\lambda},\hat{\bar{\lambda}})$ by definition parametrize the group manifold SL$(2,\mathbb{C})$. 
Explicitly, using the spinor notation for $\Lambda$ in \eqref{eq:Lambda-spinor}, 
we write the map as 
\begin{align}
    (\Lambda_{\a\da})^I{}_J = (\sigma^\m)_{\a\da}  \Lambda_\m{}^A (\sigma_A)^I{}_J = 2 \hat{\lambda}_{\a}{}^{I} \hat{\rambda}_{J\da} \,.
    \label{eq:map-body}
\end{align}
The normalization condition is recast as 
\begin{align}
    -\frac{1}{2} \e^{\a\b} \be^{\da\db} \L^{IJ}_{\a\da} \L_{KL,\b\db}
    = -2 \delta{}^{(I}{}_{K} \delta^{J)}{}_{L}
    \,.
\end{align}
The idea of exchanging the frame vectors $\Lambda^\mu{}_A$ with a pair of spinors is by no means new. 
For an early example, the Newman-Penrose dyads \cite{Newman:1961qr,Penrose:1987uia}
replaced the standard tetrads of general relativity by spinors. 
In the context of a (super)particle action, a spinor decomposition that are 
manifestly covariant with respect to both SO$(1,3)$ groups can be found  
{\it e.g.} in \cite{Delduc:1991ir,Galperin:1991gk}.

\paragraph{Position and spin} 
Having mapped the ``base" part of $T^*(\mathrm{SO}(1,3))$ between the two models, we move on to the ``fiber" part. 
We can try to find the expressions for $(x^\m, S_{\m\n})$ directly via the incidence relation 
of the twistor model, or by going through the body frame vectors $(V_a, W_a)$ at an intermediate stage. We explain both approaches,  
one after another.

In the massless twistor theory, it is well known that 
the incidence relation with complexified Minkowski coordinates conveniently account for the helicity of the particle \cite{Penrose:1972ia}.
For both massive and massless particles, the idea of interpreting spin as a displacement from the real Minkowski space 
to a complex Minkowski space 
was lucidly spelled out by Newman and Winicour \cite{Newman:1974fr}.
The idea played a key role in Shirafuji's model \cite{Shirafuji:1983zd} for massless particles with spin. 
The same idea can be incorporated into the dynamics of massive twistors, 
provided that the internal SU$(2)$ of the twistor space 
is identified with the little group of the massive particle. 

The incidence relation in our convention is given by 
\begin{align}
        \mu^{\da I} = z^{\dot{\alpha}\beta} \lambda_{\beta}{}^I \,.
        \label{eq:incidence-massive}
\end{align} 
Since $\lambda_\a{}^I$ is an invertible matrix, unlike in the massless case, 
we can explicitly solve the relation for $z^{\da\b}$. Taking the real and imaginary parts of $z^\m = x^\m  + i y^\m$, 
we obtain
\begin{align}
    \begin{split}
    x^{\da\a} &= \frac{1}{2}  \left( 
        -\frac{1}{\det(\lambda)}  \mu^{\da I} \lambda^{\a}{}_I
        +\frac{1}{\det(\bar{\lambda})} \bar{\lambda}^{I\da} \bar{\mu}_I{}^{\a} 
    \right) \,,
    \\
    y^{\da\a} &= \frac{1}{2i}  \left( 
        -\frac{1}{\det(\lambda)}  \mu^{\da I} \lambda^{\a}{}_I
        -\frac{1}{\det(\bar{\lambda})} \bar{\lambda}^{I\da} \bar{\mu}_I{}^{\a} 
    \right) \,.
    \end{split}
    \label{eq:comp-minkowski}
\end{align}
The $y^\mu$ coordinate does not appear explicitly in the spherical top model. 
Instead, it contributes to the spin variable through the relation, 
\begin{align}
    \label{eq:Spin-*yp}
    S_{\m\n} = *(y_\m p_\n - y_\n p_\m) = \ve_{\m\n}{}^{\r\s} y_\r p_\s \,.
\end{align}
When translated to the spinor notation, it gives 
\begin{align}
    \label{map-S}
    S_{\a\b} = \frac{1}{2} \left( 
        \bar{\mu}_{I(\a} \lambda_{\b)}{}^I 
        -
        \frac{1}{\det(\lambda)} \lambda_{(\a}{}^I \lambda_{\b)}{}^J (\rmu^{\dc}{}_I \rambda_{J\dc})
    \right)
    \,.
\end{align}
The orbital angular momentum in the spinor notation gives 
\begin{align}
    \label{map-L}
    L_{\a\b} = \frac{1}{2} \left( 
        \bar{\mu}_{I(\a} \lambda_{\b)}{}^I 
        +
        \frac{1}{\det(\lambda)} \lambda_{(\a}{}^I \lambda_{\b)}{}^J (\rmu^{\dc}{}_I \rambda_{J\dc})
    \right)
    \,.
\end{align}
The orbital and spin parts of the angular momentum add up to $J_{\a\b}$ in \eqref{eq:conf-twistor-massive} as expected. 
In summary, the map from the massive twistor space to the phase space of the spherical top model is 
\begin{align}
    \begin{split}
    p_{\a\da} = - \lambda_\a{}^I \bar{\lambda}_{I \da} \,, 
    &\quad
    x^{\da\a} = \frac{1}{2}  \left( 
        -\frac{1}{\det(\lambda)}  \mu^{\da I} \lambda^{\a}{}_I
        +\frac{1}{\det(\bar{\lambda})} \bar{\lambda}^{I\da} \bar{\mu}_I{}^{\a} 
    \right)  \,,
    \\
    (\Lambda_{\alpha\dot{\alpha}})^I{}_J = \frac{2 \lambda_\alpha{}^{I} \bar{\l}_{J\da} }{|\det(\lambda)|}  \,,
    &\quad 
    S_{\a\b} = \frac{1}{2} \left( 
        \bar{\mu}_{I(\a} \lambda_{\b)}{}^I 
        -
        \frac{1}{\det(\lambda)} \lambda_{(\a}{}^I \lambda_{\b)}{}^J (\rmu^{\dc}{}_I \rambda_{J\dc})
    \right)
    \,.
    \end{split}
    \label{map-all-final}
\end{align}

\paragraph{Minimal coordinates}
In the previous section, we combined $V_a$ and $W_a$ into $J_a$, $\bar{J}_a$  
and then translated them into the spinor notation. The resulting Dirac brackets were summarized in 
\eqref{eq:Dirac-LJ-SU2}:
\begin{align}
    \begin{split}
        \{ J^I{}_J, J^K{}_L \}_* &= -i ( \delta^I{}_L J^K{}_J - \delta^K{}_J J^I{}_L )
        \,,\\
        \{ (\Lambda_{\a\da})^I{}_J , J_{KL} \}_* &= + i (\L_{\a\da})^I{}_{(K}\epsilon_{L)J} 
        \,.
    \end{split} 
    \nonumber
\end{align}
We are interested in the twistor expressions for $J$, $\bar{J}$.

Just as we used  $\Lambda^\m{}_A$ to switch between the space-time frame and the body frame 
in the spherical top model, 
we may use $(\lambda,\bar{\lambda})$ to switch 
between an SL$(2,\mathbb{C})$ index and the internal SU$(2)$ index. 
We may call the SU$(2)$-indexed variables ``body frame components''.

Converting the space-time expression for $J_{\a\b}$ in \eqref{eq:conf-twistor-massive} to the body frame, 
we find 
\begin{equation}
    \label{eq:J-twistor}
    \bar{J}_{\da\db} = i\mem \frac{1}{\det(\rambda)} \rambda^I{}_{\da} \rambda^J{}_{\db} J_{(IJ)}
    \,,\quad
    J_{\a\b} = i\mem \frac{1}{\det(\lambda)} \lambda_{\a}{}^I \lambda_{\b}{}^J \bar{J}_{(IJ)}
    \,,
\end{equation}
where the body frame components are
\begin{align}
    \label{eq:JIJ-def}
    \begin{split}
        2iJ^I{}_J 
        = -2 \rmu^{\da I} \rambda_{J\da}
        = V^I{}_J + iW^I{}_J
        \,,\\
        -2i\bar{J}^J{}_I 
        = -2 \lmu_I{}^{\a}{} \lambda_{\a}{}^J
        = V^I{}_J - iW^J{}_I
        \,.
    \end{split}
\end{align}
In fact, these expressions contain not only the 
$\mathfrak{su}(2)$ part but also the $\mathfrak{u}(1)$ part:
\begin{align}
    V_{(IJ)} = -\left(
        \rmu^{\da}{}_{(I} \rambda_{J)\da} +\lambda_\a{}_{(I} \lmu_{J)}{}^\a
    \right)
    &\,,\quad
    V_0 = \frac{1}{2}\left(
        [\rambda\rmu] +\langle\lmu\lambda\rangle
    \right) = D
    \,,\\
    W_{(IJ)} = i\left(
        \rmu^{\da}{}_{(I} \rambda_{J)\da} -\lambda_\a{}_{(I} \lmu_{J)}{}^\a 
    \right)
    &\,,\quad
    W_0 = \frac{1}{2i}\left(
        [\rambda\rmu] -\langle\lmu\lambda\rangle
    \right) 
    \,.
    \label{eq:W-twistor}
\end{align}
To see the fate of the $\mathfrak{u}(1)$ part, we note that 
\begin{equation}
    \label{eq:VW-xy}
    V^A = -|\det(\lambda)| (x\cdot\L^A)
    \,,\quad
    W^A = -|\det(\lambda)| (y\cdot\L^A)
    \,,
\end{equation}
where $\Lambda$ should be understood as in \eqref{eq:map-body} and $x$ and $y$ as in \eqref{eq:comp-minkowski}. 
These are the twistor counterpart of \eqref{eq:VW-body}. 
Since $V^0=0=W^0$ in the spherical top model, we should require the same 
in the twistor model. The requirement is fulfilled by the gauge-fixing conditions 
$\chi$ in \eqref{spinor-constraint} which 
is equal to $-(V^0+iW^0)/\det(\bar{\lambda})$, or 
the time-like body frame component of $z^\m = x^\m + i y^\m$.

In summary, we have shown that the vectors $V$ and $W$ in the two models coincide perfectly 
on the physical subspace. We have also reconfirmed 
the identification of the internal SU$(2)$ as the little group; see \eqref{su2-twistor}.
Separating the orbital and spin parts of \eqref{eq:J-twistor}, 
we can rewrite \eqref{map-S} and \eqref{map-L} as 
\begin{equation}
    \label{eq:LS-rederived}
    L_{\a\b} = -\frac{1}{2\mem\det(\lambda)} \lambda_{\a}{}^I \lambda_{\b}{}^J V_{(IJ)}
    \,,\quad
    S_{\a\b} = i  \frac{1}{2\mem\det(\lambda)} \lambda_{\a}{}^I \lambda_{\b}{}^J W_{(IJ)}
    \,.
\end{equation}
It implies the transverse spin-squared $\tilde{S}^2$ appearing in \eqref{spinor-constraint} can be expressed as 
\begin{align}
    \tilde{S}^2 = -\frac{1}{2} W^{(IJ)}W_{(IJ)} = W^a W_a
    \,.
    \label{eq:S2-twistor}
\end{align}

\subsection{Physical phase space} 

\paragraph{Dirac brackets}

Our goal is to check whether the massive twistor phase space equipped with the constraints \eqref{spinor-constraint} reproduce the Dirac bracket \eqref{Dirac-spinor}.
The Dirac bracket is constructed in the usual manner:
\begin{align}
\begin{split}
    \{ f, g \}_* = \{ f , g \} 
    &-\big( \{f , \phi \} \{ \bar{\chi} , g \} - \{f , \bar{\chi} \} \{ \phi , g \} \big)
    \\
    &-\big( \{f , \bar{\phi} \} \{ \chi , g \} - \{f , \chi \} \{ \bar{\phi} , g \} \big)
    \,.
\end{split}
\end{align}
On the constraint surface, the non-vanishing terms are 
\begin{align}
    \nonumber
    \{\lambda_\a{}^I, \lambda_\b{}^J\}_*
    &= -\frac{m'}{2im}\left[
        \big(\lambda_\a{}^K{W}_K{}^I\big) \lambda_\b{}^J
        - \lambda_\a{}^I \big(\lambda_\b{}^K{W}_K{}^J\big)
    \right]
    \,,\\
    \nonumber
    \{\lambda_\a{}^I, \rambda_{J\db}\}_*
    &= -\frac{m'}{2im}\left[
        \big(\lambda_\a{}^K{(W}_K{}^I\big) \rambda_{J\db}
        + \lambda_\a{}^I \big({W}_J{}^K\rambda_{K\db}\big)
    \right]
    \,,\\
    \nonumber
    \{\lambda_\a{}^I, \lmu_J{}^\b\}_*
    &= \left(
        \delta_\a{}^\b \delta^I{}_J + \frac{1}{2m} \lambda_\a{}^I \lambda^\b{}_J 
    \right)
    + \frac{m'}{2im}\left[
        \big(\lambda_\a{}^K{W}_K{}^I\big) \lmu_J{}^\b
        - \lambda_\a{}^I \big({W}_J{}^K\lmu_K{}^\b\big)
    \right]
    \,,\\
    \nonumber
    \{\lambda_\a{}^I, \rmu^{\db J}\}_*
    &= \frac{m'}{2im}\left[
        \big(\lambda_\a{}^K{W}_K{}^I\big)\rmu^{\db J}
        + \lambda_\a{}^I\big(\rmu^{\db K}{W}_K{}^J\big)
    \right]
    \,,\\
    \nonumber
    \{\rmu^{\da I}, \rmu^{\db J}\}_*
    &= 
    \frac{1}{2m} \left(
        \rmu^{\da I}\rambda^{J\db} - \rmu^{\db J}\rambda^{I\da}
    \right)
    + \frac{m'}{2im}\left[
        \big(\mu^{\da K}{W}_K{}^I\big)\rmu^{\db J} - \rmu^{\da I}\big(\rmu^{\db K}{W}_K{}^I\big)
    \right]
    \,,\\
    \label{eq:Dirac-twistor}
    \{\rmu^{\da I}, \lmu_J{}^\b\}_*
    &= \frac{m'}{2im}\left[
        \big(\rmu^{\da K}{W}_K{}^I\big)\lmu_J{}^\b
        + \rmu^{\da I}\big({W}_J{}^K\lmu_K{}^\b\big)
    \right]
    \,,
\end{align}
and their complex conjugates.
The algebra may look unwieldy, but its actual content is rather simple. 
If we focus on the normalized variables defined in \eqref{eq:twistor-normalized}, 
\begin{align}
\begin{split}
    \hat{\lambda}_\a{}^I := (\det(\lambda))^{-1/2} \lambda_\a{}^I
    &\,,\quad
    \hat{\rmu}^{\da I} := (\det(\lambda))^{1/2} \rmu^{\da I}
    \,,\\
    \hat{\rambda}_{I\da} := (\det(\rambda))^{-1/2} \rambda_{I\da}
    &\,,\quad
    \hat{\lmu}_I{}^\a := (\det(\rambda))^{1/2} \lmu_I{}^\a
    \,,\\
\end{split}
\nonumber 
\end{align}
the ``Regge slope'' $m'(\tilde{S}^2)$ terms disappear completely from the Dirac brackets:
\begin{align}
\label{eq:Dirac-lm-normalized}
\begin{split}
    &\{\hat{\lambda}_\a{}^I, \hat{\lambda}_\b{}^J\}_* = 0
    \,,\quad
    \{\hat{\lambda}_\a{}^I, \hat{\rambda}_{J\db}\}_* = 0
    \,,\\
    &\{\hat{\lambda}_\a{}^I, \hat{\lmu}_J{}^\b\}_* 
    = \delta_\a{}^\b\delta^I{}_J + {\textstyle\frac{1}{2}}\,\hat{\lambda}_\a{}^I\hat{\lambda}^\b{}_J
    \,,\quad
    \{\hat{\lambda}_\a{}^I, \hat{\rmu}^{\db J}\}_* 
    = 0
    \,,\\
    &\{\hat{\rmu}^{\da I}, \hat{\rmu}^{\db J}\}_*
    = \frac{1}{2} \left(
        \hat{\rmu}^{\da I}\hat{\rambda}^{J\db} - \hat{\rmu}^{\db J}\hat{\rambda}^{I\da}
    \right)
    \,,\quad
    \{\hat{\rmu}^{\da I}, \hat{\lmu}_J{}^{\b}\}_*
    = 0
    \,.
\end{split}
\end{align}
This observation is consistent with the fact that the Dirac bracket algebra of $(\L_{\a\da})^I{}_J$ and $J^I{}_J$ 
is isomorphic to the Poison bracket algebra of $T^*(\mathrm{SO}(1,3))$ 
and is independent of the Regge trajectory. 
Indeed, expressing $(\L_{\a\da})^I{}_J$ and $J^I{}_J$ in terms of the normalized twistor variables,
we reproduce the fully constrained Dirac brackets of the spherical top model in the minimal coordinates in \eqref{eq:Dirac-LJ-SU2}.

In switching back to the original coordinates, 
the only non-trivial step is to multiply each variable by appropriate factors of $\det(\lambda) = m(W^2)$. 
Using the explicit map \eqref{map-all-final}, 
we can eventually reproduce the Dirac bracket algebra of the spherical top model \eqref{Dirac-spinor}.

\subsection{Equation of motion} \label{dynamics-new}
\paragraph{First-order action}
A simple and natural choice of the symplectic potential on the twistor space, 
respecting all the symmetries of the space, is given by
\begin{align}
    \begin{split}
    \theta &= \frac{i}{2} \left(
        \bar{Z}_I{}^\rmA d Z_\rmA{}^I - Z_\rmA{}^I d \bar{Z}_I{}^\rmA
    \right)
    \\
    &= 
    - \lambda_\a{}^I \rambda_{I\da} dx^{\da\a} -iy^{\da\a} \left(
        d\lambda_{\a}{}^I \rambda_{I\da} - \lambda_{\a}{}^I d\rambda_{I\da}
    \right)
    \,,
    \end{split}
\end{align}
It is a straightforward two-twistor generalization of Shirafuji's action~\cite{Shirafuji:1983zd} for a massless twistor.
In the second line, $x^{\da\a}$ and $y^{\da\a}$ should be understood as functions of 
twistor variables as in \eqref{eq:comp-minkowski}.
Equivalently, $\theta$ here may be regarded as the pull-back of the symplectic 
potential~\eqref{symp-1} of the spherical top model 
through the map~\eqref{map-all-final}. 

The first-order action is
\begin{equation}
    \label{eq:action-twistor}
    \mathcal{S} = \int d\sigma\left(
        \frac{i}{2}\big(\bar{Z}_I{}^\rmA \dot{Z}_\rmA{}^I - Z_\rmA{}^I \dot{\bar{Z}}_I{}^\rmA\big)
        - \bar{\kappa} \phi - \kappa \bar{\phi}
    \right)
    \,,
\end{equation}
where $\phi$ and $\bar{\phi}$ are the mass-shell constraints in \eqref{spinor-constraint}. 
Although the symplectic potential respects conformal symmetry, the gauge generators break the dilatation symmetry 
(this can be made explicit by introducing infinity twistors).
The variations of the action give
\begin{equation}
\label{eq:eom-lm}
\begin{split}
    \delta_{\lmu} \mathcal{S} = 0 &\,\,\implies\,\, %
    \dot{\lambda}_{\a}{}^I = -i(\Re\,\kappa) m'  \lambda_{\a}{}^J W_J{}^I
    \,,
    \\
    \delta_{\rambda} \mathcal{S} = 0 &\,\,\implies\,\, %
    \dot{\rmu}^{\da I} = -i(\Re\,\kappa) m' \rmu^{\da}{}^J W_J{}^I + \frac{\kappa}{2}\mem \rambda{}^{\da I}
    \,,
\end{split}
\end{equation}
and their complex conjugates.

\paragraph{Lagrange multipliers}
Requiring that the gauge-fixing condition $W^0 = 0$ is preserved by the equation of motion, 
we find that the Lagrange multipliers are subject to the condition, 
\begin{align}
    0= 2\dot{W}^0
    = \Im\big[\kappa\mem\det(\rambda)\big]
    \,.
\end{align}
Then, $\phi=0$ and $\bar{\phi}=0$ imply that $(\Im\,\kappa) = 0$.
Rather than imposing another gauge-fixing condition, 
we solve for $(\Re\,\kappa)$ from the resulting equations of motion as we did in section~\ref{sec:vect-eom}.
From the complexified incidence relation \eqref{eq:incidence-massive},
\begin{align}
    \dot{z}^{\da\a}
    = -\frac{\kappa}{2m} p^{\da\a}
    \,\,\implies\,\,
    \dot{x}^{\m} = (\Re\,\kappa)\frac{1}{m} p^{\m}
    \,,\quad
    \dot{y}^{\m} = 0
    \,.
\end{align}
Hence,
\begin{align}
    (\Re\,\kappa) = 
    \frac{d\tau}{d\sigma}
    \,,\quad
    d\tau = \sqrt{-dx_\m\mem dx^\m}
    \,.
\end{align}
Additional calculations based on the map \eqref{map-all-final} give the final equations of motion:
\begin{align}
    \label{eq:eom-all-tau-fromtwistor}
    p^\m = m \frac{dx^\m}{d\tau}
    \,,\quad
    \frac{dp_\m}{d\tau} = 0
    \,,\quad
    \frac{d\L_\m{}^a}{d\tau} = - \ve^a{}_{bc} (2m' W^b) \L_\m{}^c
    \,,\quad
    \frac{dW_a}{d\tau} = 0
    \,.
\end{align}
The particle propagates with constant velocity and spin angular momentum, while the body frame rotates with a constant three-angular velocity $\omega^a := 2m'W^a$.
The minus sign of $(-\ve^a{}_{bc})$ accounts correctly for the fact that 
the angular velocity $\omega^a$ is measured with respect to the body frame.
It is easy to verify that \eqref{eq:eom-all-tau-fromtwistor} conforms to the result \eqref{eq:eom-all-tau} 
of section~\ref{sec:vect-eom}.

In terms of the constant of motion $\omega^a$, \eqref{eq:eom-lm} can be integrated to give 
\begin{align}
    \lambda_\a{}^I(\tau) = \lambda_\a{}^J(0)\mem U_J{}^I(\tau)
    \,,\quad
    \mu^{\da I}(\tau) = \frac{\tau}{2}\mem \rambda^{I\da}(\tau) + \mu^{\da J}(0)\mem U_J{}^I(\tau)
    \,,
\end{align}
where $U(\tau) = \exp[\mem\omega^a\tau\,(\sigma_a)/2i\mem]$.
This translates to
\begin{align}
\begin{split}
    p_\m(\tau) = p_\m(0)
    &\,,\quad
    \L_\m{}^a(\tau) = \mem (U^{-1})^a{}_b(\tau) \mem \L_\m{}^b(0)
    \,,\\
    x^\m(\tau) =
    x^\m(0) + \L^\m{}_0(0) \mem\tau
    &\,,\quad
    S_{\m\n}(\tau) = S_{\m\n}(0)
    \,,
\end{split}
\end{align}
where $(U^{-1} \s^a U)_I{}^J := U^a{}_b (\s^b)_I{}^J$.
The position and the spin can be conveniently extracted from the $J^I{}_J$ matrix \eqref{eq:JIJ-def} 
or the complexified position \eqref{eq:comp-minkowski}.

\subsection{Quantization} 

We review the quantization of the massive twistor model \cite{Fedoruk:2014vqa} in our notation. 
We quantize the unconstrained phase space first and 
then impose the constraints. 
The quantization is straightforward in the $(\lambda,\bar{\lambda})$ polarization, 
where the conjugate variables act as derivatives:
\begin{align}
    \bar{\mu}_I{}^\a = -i \frac{\partial}{\partial \lambda_\a{}^I} \,,
    \quad 
    \mu^{\da I} = -i \frac{\partial}{\partial \bar{\lambda}_{I\da} } \,. 
    \label{eq:mu-lambda-derivative}
\end{align}
Ignoring the SL$(2,\mathbb{C})$ indices for a moment and focusing on the SU$(2)$ indices, 
for an integer or half-integer $s$, 
any totally symmetrized polynomial $P^{(2s)}(\lambda, \bar{\lambda})$ of degree $2s$ 
is an eigenstate of the transverse spin-squared: 
\begin{align}
    \tilde{S}^2 P^{(2s)}(\lambda, \bar{\lambda}) = s(s+1) P^{(2s)}(\lambda, \bar{\lambda}) \,, 
\end{align}
where $\tilde{S}^2$ should be understood as a differential operator through 
\eqref{eq:W-twistor}, \eqref{eq:S2-twistor} and \eqref{eq:mu-lambda-derivative}.
The mass spectrum of the particle follows from 
the Regge trajectory as 
\begin{align}
    (-p^2) P^{(2s)}(\lambda,\rambda) = m^2(\tilde{S}^2) P^{(2s)}(\lambda,\rambda) 
    = m^2(s(s+1)) P^{(2s)}(\lambda,\rambda) \,.
\end{align}
In the quantum theory, once we impose the constraints $\phi = 0 = \bar{\phi}$, 
the wave-function is forced to be independent of the conjugate variables 
$(\bar{\chi},\chi)$, 
just as $p=0$ for a classical trajectory translates to 
$\partial_x \psi(x) = 0$ for the corresponding quantum wave-function.

\paragraph{On-shell fields}
Restoring the SL$(2,\mathbb{C})$ indices, we may distinguish $(2s+1)$ types of polynomials, 
each of which can be mapped to an on-shell space-time field \cite{Fedoruk:2014vqa} as 
\begin{align}
    \begin{split}
        \Phi^{(2s,0)}_{\a_1 \a_2 \ldots \a_{2s}} (x) 
        &=  c_{I_1 \ldots I_{2s}} \int d[\lambda,\bar{\lambda}] \, 
            \lambda_{\a_1}^{I_1}  
            \lambda_{\a_2}^{I_2} \cdots 
            \lambda_{\a_{2s}}^{I_{2s}} \,
        \exp\left( -i  x^{\db\b} \lambda_\b{}^J \bar{\lambda}_{J\db}  \right) \,,
        \\   
        \Phi^{(2s-1,1)}_{\da_1 | \a_2 \ldots \a_{2s}} (x)  
        &= c_{I_1 \ldots I_{2s}} \int d[\lambda,\bar{\lambda}] \, 
            \bar{\lambda}^{I_1}_{\da_1}  
            \lambda_{\a_2}^{I_2} \cdots 
            \lambda_{\a_{2s}}^{I_{2s}} \,
        \exp\left( -i  x^{\db\b} \lambda_\b{}^J \bar{\lambda}_{J\db}  \right) \,,
        \\
        \vdots \qquad\,\,\, & \qquad\,\,\, \vdots
        \\
        \Phi^{(0,2s)}_{\da_1 \da_2 \ldots \da_{2s}} (x)  
        &= c_{I_1 \ldots I_{2s}} \int d[\lambda,\bar{\lambda}] \, 
            \bar{\lambda}^{I_1}_{\da_1}  
            \bar{\lambda}^{I_2}_{\da_2} \cdots \bar{\lambda}^{I_{2s}}_{\da_{2s}} \,
        \exp\left( -i  x^{\db\b} \lambda_\b{}^J \bar{\lambda}_{J\db}  \right) \,,
    \end{split}
    \label{eq:quantum-spectrum}
\end{align}
where $c_{I_1 \ldots I_{2s}}$ is a totally symmetrized SU$(2)$ tensor, and the integration measure is 
\begin{align}
    d[\lambda,\bar{\lambda}] = {d^4\lambda}\mem {d^4\bar{\lambda}} 
    \left[ \delta(\det(\lambda) - m) \delta(\det(\bar{\lambda}) - m) \right] \,.
\end{align} 
The on-shell fields satisfy the coupled, first-order, Dirac-Fierz-Pauli-Bargmann-Wigner equations, 
which implies that each field satisfies the Klein-Gordon equation.
Each line in \eqref{eq:quantum-spectrum} shares the same SU(2) tensor $c_{I_1 \ldots I_{2s}}$ 
and represents the same set of $(2s+1)$ physical states, 
although a naive counting of all possible $P^{(2s)}(\lambda, \bar{\lambda})$'s gives a much larger number.

We close this subsection with a few comments.
First, we note that the spectrum of the quantum states shown in \eqref{eq:quantum-spectrum} 
takes precisely the same form as the one based on massive spinor-helicity variables of 
Arkani-Hamed, Huang and Huang \cite{Arkani-Hamed:2017jhn}. 
Second, 
it is well known that a twistor description 
reveals many interesting properties of scattering amplitudes of massless particles 
\cite{Witten:2003nn}. 
It would be interesting to see, to what extent, the twistor point of view would become 
useful in understanding scattering amplitudes of massive particles. 
Technically, 
the ``half-Fourier-transform" from the $(\lambda,\bar{\lambda})$ polarization 
to the twistor $(\lambda,\mu)$ polarization may face complications 
due to the mass-shell constraint.

\section{Discussion \label{sec:discussion} }

We have established the equivalence between the spherical top model and the massive twistor model. 
Undoubtedly, this is only the beginning of a longer story. 

An immediate question is whether the equivalence persists in an interacting theory.
We could try to couple the spinning particle to an external electromagnetic field 
and see if we recover the Bargmann-Michel-Telegdi (BMT) equation \cite{Bargmann:1959gz},  
or try to couple it to an external gravitational field 
and see if we recover the Mathisson–Papapetrou–Dixon (MPD) equation \cite{Mathisson:1937zz,Papapetrou:1951pa,Dixon:1970zza}.
Coupling to an external field may not be a straightforward exercise 
as it requires generalizing the spin-gauge constraints in a way 
consistent with the U(1) gauge invariance of electromagnetism 
or the general covariance of gravity.

For the electromagnetic coupling, the internal U(1) of the twistor model can be associated with the electric charge \cite{Newman:1974fr}.
An intriguing application of this idea is Bette's derivation of a ``square-root'' of the BMT equation \cite{Bette:1989zt}. 
It would be interesting to identify the spherical top counterpart of the equation 
and confirm the classical value of the $g$-factor: $g=2$. 

While the twistor model conforms to the covariant spin-gauge by construction, 
the spherical top model allows for other gauge choices. 
We may keep the gauge generators $\phi_a$ but use gauge-fixing conditions different from $\chi^a$ used in this paper. 
For instance, when describing a binary system of two spinning particles weakly interacting through gravity, 
we may choose a gauge in which the center-of-momentum frame plays a prominent role. 

A version of the spherical top model constitutes the free part of an 
effective field theory (EFT) approach \cite{Goldberger:2004jt} to the post-Newtonian (PN) gravity of spinning objects \cite{Levi:2015msa,Levi:2017kzq}. 
In intermediate steps of the EFT computation, 
some higher time-derivative terms appear in the Lagrangian, 
but they can be eliminated systematically in the final Hamiltonian of the interacting binary system. 
It would be interesting to experiment with different gauge choices and see if one can reduce the appearance of higher derivative terms. 

Recently, some spin-dependent terms of the EFT Hamiltonian have been 
compared with a post-Minkowskian (PM) Hamiltonian derived from a method based on scattering amplitudes \cite{Chung:2018kqs,Chung:2020rrz} 
in the overlapping regime of validity.
The comparison of the final results required non-trivial canonical transformation between 
phase space coordinates in the two computational frameworks. 
Since we showed that the spherical top model is equivalent to the massive twistor model, 
and that the latter upon quantization reproduces the on-shell states in terms of the massive spinor-helicity variables,
a closer look at the gauge choices of the spherical top model 
may reveal a more direct connection between the EFT-based PN Hamiltonian and the amplitude-based PM Hamiltonian. 
This possibility was one of the main motivations for the authors to examine the spherical top model in the first place.

Ever since the work of Witten \cite{Witten:2003nn}, theories based on (or inspired by) massless twistors have made tremendous contributions to scattering amplitudes of massless particles. It is tempting to conjecture that 
a similar degree of success with massive twistors is on the horizon.

\acknowledgments
We are grateful to Yu-tin Huang, Mich\`ele Levi, Rodolfo Russo, Congkao Wen, Chris White 
for discussions and comments on the manuscript.
The work of JHK and SL was supported in part by the National Research Foundation of Korea grant NRF-2019R1A2C2084608. JWK was supported by the Science and Technology Facilities Council (STFC) Consolidated Grant ST/T000686/1 \emph{``Amplitudes, Strings and Duality”}.

\newpage 
\appendix 

\section{Hamiltonian mechanics under constraints}
\label{sec:constrained} 

We give a minimal review of the Hamiltonian mechanics of constrained systems. 
For more comprehensive expositions, see {\it e.g.} \cite{Dirac:text,Henneaux:1992ig}.

\paragraph{Basic conventions}

Here are our conventions for symplectic form, Poisson bracket, etc. 
For a flat phase space $\mathbb{R}^{2n}$ with coordinates $(q^i , p_i)$, 
we orient the symplectic form $\omega$ as 
\begin{align}
\omega = dp_i \wedge dq^i \,.
\end{align}
When a Hamiltonian $H$ generates a time flow $V=d/dt$, 
Hamilton's equation reads 
\begin{align}
dH = - i_V \omega \,.
\label{Hamiltonian-eq1}
\end{align}
Short computations,
\begin{align}
dH = \left(\frac{\partial H}{\partial q^i} \right) dq^i + \left(\frac{\partial H}{\partial p_i} \right) dp_i \,, 
\quad 
V = \frac{d}{dt} = \dot{q}^i \frac{\partial}{\partial q^i} + \dot{p}_i \frac{\partial}{\partial p_i} \,,
\end{align}
verify that the geometric form of the equation \eqref{Hamiltonian-eq1} agrees with the conventional form, 
\begin{align}
\dot{p_i} = - \left(\frac{\partial H}{\partial q^i} \right)  \,, \quad
\dot{q}^i =  \left(\frac{\partial H}{\partial p_i} \right) \,. 
\end{align}
Geometrically, the Poisson bracket is defined as
\begin{align}
\{ f , g \} = -\omega(X_f, X_g) \,, 
\;\;
\mbox{where}
\;\; 
df = - i_{X_f} \omega \,,
\;\;
dg = - i_{X_g} \omega 
\quad 
\Longrightarrow
\quad
\frac{df}{dt} = \{ f, H \} \,.
\end{align}
For $\mathbb{R}^{2n}(q^i , p_i)$, since 
\begin{align}
df = \left(\frac{\partial f}{\partial q^i} \right) dq^i + \left(\frac{\partial f}{\partial p_i} \right) dp_i \,, 
\quad 
X_f = -\left(\frac{\partial f}{\partial q^i} \right) \frac{\partial}{\partial p_i} + \left(\frac{\partial f}{\partial p_i} \right) \frac{\partial}{\partial q^i} \,,
\end{align}
and similarly for $g$, the bracket computes
\begin{align}
\{ f, g \} = \frac{\partial f}{\partial q^i}\frac{\partial g}{\partial p_i} - \frac{\partial g}{\partial q^i}\frac{\partial f}{\partial p_i} \,.
\label{Poisson-explicit}
\end{align}
Equivalently, we can begin with specifying the elementary Poisson bracket, 
\begin{align}
\{ q^i , q^j \} = 0 = \{p_i , p_j\} 
\,,
\quad 
\{ q^i , p_j \} = \delta^i{}_j \,,
\label{Poisson-elementary}
\end{align}
and require antisymmetry, Leibniz rule and Jacobi identity to recover \eqref{Poisson-explicit}.

Our conventions can be used for general symplectic manifolds with little modifications: 
\begin{align}
\omega = \frac{1}{2} \omega_{mn} dy^m \wedge dy^n \,,
\quad 
\{ f, g \} 
= \omega^{mn} \frac{\partial f}{\partial y^m} \frac{\partial g}{\partial y^n}  \,, 
\quad 
\omega^{mk}\omega_{kn} = \delta^m{}_n \,.
\end{align}
When momentum and position can be distinguished, we orient the symplectic form such that the momentum precedes the position. 
It is often convenient to introduce the symplectic potential satisfying $\omega = d\theta$.  
While $\omega$ is required to be globally well-defined, $\theta$ may be only locally well-defined.

\paragraph{Constrained system and Dirac bracket}

We give a brief review of the Hamiltonian mechanics of a constrained system, focusing on the case when the constraints satisfy a few simplifying properties. We start with $n$ position-momentum pairs and let the constraints remove $k$ pairs. The constraints are divided into `gauge generators' $\phi_A$ and `gauge-fixing conditions' $\chi^A$ $(A=1,\ldots,k)$. We require that their  Poisson brackets take the form 
\begin{align}
\{ \phi_A , \phi_B \} \approx 0 \approx \{ \chi^A , \chi^B \} \,,
\quad 
C^A{}_B : = \{ \chi^A , \phi_B \} \,, 
\quad 
\det(C) \neq 0 \,.
\end{align}
The weak equality `$\approx$' means that the equality holds on the constraint surface.
The gauge generators among themselves are said to form first-class constraints. The gauge generators and the gauge-fixing conditions together are said to form second-class constraints. Finally, 
to qualify as `symmetry' generators, $\phi_A$ should commute with the Hamiltonian $H$. 

These constraints provide an ideal setup to introduce the Dirac bracket. We motivate the Dirac bracket starting from the Hamiltonian action principle. Consider the action 
\begin{align}
\mathcal{S} = \int dt \left( p_i \dot{q}^i - H - \kappa^A \phi_A \right) \,.
\end{align}
The linear combination of constraints, $\kappa^A \phi_A$, serves two purposes. First, it imposes the constraints $\phi_A = 0$ through the Lagrange multipliers $\kappa_A$: 
\begin{align}
\delta_\kappa \mathcal{S} = 0 
\quad 
\Longrightarrow 
\quad 
\phi_A = 0 \,.
\end{align}
Second, it generates a flow along the gauge orbits to accompany the Hamiltonian time flow:
\begin{align}
\delta_{p,q} \mathcal{S} = 0 
\quad 
\Longrightarrow 
\quad 
\frac{df}{dt} = \{ f ,H \} + \kappa^A \{ f, \phi_A \} \,.
\label{time-gauge}
\end{align}
The two types of equations are mutually consistent since 
\begin{align}
0 = \frac{d\phi_A}{dt} = \{ \phi_A ,H \} + \kappa^B \{ \phi_A , \phi_B \} \approx 0 \,.
\end{align}

Now, if we impose the conditions $\chi^A =0$, the multipliers $\kappa^A$ should be tuned such that the trajectory does not deviate from the gauge slice. Consistency of the procedure requires
\begin{align}
\begin{split}
0 &= \frac{d\chi^A}{dt} = \{ \chi^A ,H \} + \kappa^B \{ \chi^A , \phi_B \} 
= \{ \chi^A ,H \} + \kappa^B C^A{}_B \approx 0 
\\
&\Longrightarrow
\quad 
\kappa^A = - (C^{-1})^A{}_B \{\chi^B, H\}\,.
\end{split}
\end{align}
Substituting it into \eqref{time-gauge}, we find 
\begin{align}
\begin{split}
\frac{df}{dt} &= \{ f ,H \} - \{ f, \phi_A \} (C^{-1})^A{}_B \{\chi^B, H \} 
\\
&= \{ f ,H \} - (C^{-1})^A{}_B \left( \{ f, \phi_A \}  \{\chi^B, H \} - \{ f, \chi^B \} \{\phi_A , H \} 
\right) 
\approx \{ f, H \}_* \,,
\end{split}
\end{align}
where we inserted a vanishing term ($\{\phi_A , H \} \approx 0$) to make contact with the Dirac bracket.

\paragraph{Hamiltonian mechanics on a group manifold}
In a Hamiltonian system, a continuous symmetry comes with a generator $Q$ and the associated vector flow $X_Q$ satisfying 
\begin{align}
    \{H,Q\} = 0 \,,
    \quad
    dQ = - i_{X_Q} \omega\,,
    \quad 
    X_Q[F] = \{ F , Q \} \,.
\end{align}
When there are two or more continuous symmetries, a Lie algebra structure arises: 
\begin{align}
    \{ Q_a , Q_b \} = + f^c{}_{ab} Q_c \,,
    \quad 
    [X_a , X_b] = -  f^c{}_{ab} X_c \,.
    \label{eq:structure-function}
\end{align}
The ``structure function" $f^c{}_{ab}$ should satisfy the Jacobi identity at every point. 

Now we specialize to the Hamiltonian mechanics of a particle on a group manifold. 
Let $G$ be a Lie group and $\mathfrak{g}\cong T_e(G)$ be its Lie algebra. 
The phase space is the cotangent bundle $T^*(G)$ whose fiber is $\mathfrak{g}^*$. 
The continuous isometry group of $G$ is $(G\times G)/Z$, where the two $G$ factors denote 
the left and right actions, and $Z$ denotes the center. 
We choose to use the right action to push-forward the Lie algebra of $\mathfrak{g}$ to the group manifold $G$, 
and then up-lift it to the phase space $T^*(G)$. 
The structure functions \eqref{eq:structure-function} become constants. 

We summarize a few crucial results on the extension of $\mathfrak{g}$ to $T^*(G)$. 
Let $\Theta = dg\mem g^{-1}$ be the right-invariant Maurer-Cartan one-form. 
In a suitably chosen basis, the Maurer-Cartan equation reads 
\begin{align}
    d\Theta^c = \frac{1}{2} f^c{}_{ab} \Theta^a \wedge \Theta^b \,.
\end{align}
The right-invariant symplectic potential and symplectic form are 
\begin{align}
    \theta = p_c \Theta^c \,,
    \quad 
    \omega = d\theta = d p_c \wedge \Theta^c + p_c \left( \textstyle{\frac{1}{2}} f^c{}_{ab} \Theta^a \wedge \Theta^b \right) \,.
\end{align}
Let $\Xi_a$ be the vector fields dual to the one-forms $\Theta^a$: $\Theta^a(\Xi_b) = \delta^a_b$. 
Its uplift to $T^*(G)$ is 
\begin{align}
    X_a[F] = \{ F , p_a \} =\left( \Xi_a - p_c f^c{}_{ab} \frac{\partial}{\partial p_b}  \right) F \,,
\end{align}
where $F$ is any function on $T^*(G)$.
The Lie algebra takes the same form in the two incarnations,  
\begin{align}
    [\Xi_a , \Xi_b] = -  f^c{}_{ab} \Xi_c 
    \quad 
    \mapsto
    \quad
    [X_a , X_b] = -  f^c{}_{ab} X_c \,.
\end{align}
In terms of the momenta $p_a$, the Lie algebra reads 
\begin{align}
    \{ g , p_a \} = \Xi_a g \,, 
    \quad
    \{ p_a , p_b \} = + f^c{}_{ab} p_c \,.
\end{align}
In the first equation, a group element $g$ is treated as a matrix-valued function on $G$. 

Identifying the relativistic spherical top as the right-invariant Hamiltonian system on the cotangent bundle of the Poincar\'e group uniquely determines the symplectic structure in \eqref{poisson-1} and \eqref{symp-1}%
: $\theta = \frac{1}{2}J_{\m\n}\Theta^{\m\n} + p_\m \Theta^\m$.
The right-invariant one-forms $\Theta^a$ are 
\begin{align}
    \label{eq:MCforms-Poincare}
    \begin{split}
        \Theta^{\m\n} &= -d \L^\m{}_A \mem\L^{\n A} = \L^\m{}_A \mem d\L^{\n A}
        \,,\\
        \Theta^{\m} &= dx^{\m} + \Theta^\m{}_\n x^\n
        = \L^\m{}_A \mem d( \L_\r{}^A x^\r )
        \,,
    \end{split}
\end{align}
where an element of the Poincar\'e group is described by the pair $(\Lambda^\m{}_A,x^\m)$.
As the mixed index types indicate, the ``top parametrization'' $\L^\m{}_A$ 
distinguishes the left and right actions.
Note that, if the Poincar\'e group is regarded as the semidirect product $\mathrm{SO}(1,3)\ltimes\mathbb{R}^{1,3}$, the symplectic potential can also be written as $\theta = \frac{1}{2}S_{\m\n}\Theta^{\m\n} + p_\m dx^\m$. 
The two expressions are equal, provided that $J_{\m\n} - S_{\m\n} = 2x_{[\m}p_{\n]}$. 

\newpage
\section{Spinors and twistors} 
\label{sec:spinor-conv}

\paragraph{$\mathrm{SU}(2)$ spinors} 
The $\epsilon^{IJ}$ of $\mathrm{SU}(2)$ denotes the antisymmetric rank-2 tensor 
normalized such that $\epsilon^{12} \deq {+1}$.
Its inverse is the lower-index epsilon tensor:
\begin{equation}
    \epsilon^{IK} \epsilon_{KJ} = \delta^I{}_J = \delta_J{}^I \,. 
\end{equation}
The spinor indices ($I,J,\cdots$) carry the fundamental representation of $\mathrm{SU}(2)$ and are raised and lowered by the epsilon tensors with the convention
\begin{equation}
    \psi^{I} = \epsilon^{IJ} \psi_J
    \,,\quad
    \psi_{I} = \epsilon_{IJ} \psi^J
    \,.
\end{equation}
The representation is pseudo-real; complex conjugation exchanges upper and lower indices as $[\psi_I]^* = \bar{\psi}^I$.
Note also that
\begin{equation}
    [\epsilon^{IJ}]^* = -\epsilon_{IJ}.
\end{equation}
The three Hermitian generators of $\mathrm{SU}(2)$ are half of the Pauli matrices $(\sigma_a)_{I}{}^{J}$.
The invariance of the epsilon tensors under $\mathrm{SU}(2)$ transformations implies that 
\begin{equation}
    (\sigma_a)_{IJ} = \epsilon_{JK} (\sigma_a)_{I}{}^{K}
    \,,\quad
    (\sigma_{a})^{IJ} = \epsilon^{IK} (\sigma_a)_{K}{}^{J}
\end{equation}
are symmetric in their two spinor indices.

An arbitrary spin-$s$ irreducible representation of $\mathrm{SU}(2)$ is described by a totally symmetric $\mathrm{SU}(2)$ tensor with $2s$ indices:
\begin{equation}
    \psi_{I_1 I_2 \cdots I_{2s}} = \psi_{(I_1 I_2 \cdots I_{2s})}
    \,.
\end{equation}
Specifically, for $s=1$, $(\sigma^a)_{IJ}$ and $(\sigma_a)^{IJ}$ serve as the Clebsch-Gordan coefficients. 
Consider
\begin{equation}
    (\sigma_a)^{IJ} (\sigma^a)_{KL} = -2 \delta^I_{(K} \delta^J_{L)}
    \,,\quad
    (\sigma^a)_{IJ} (\sigma_b)^{IJ}  = -2 \delta^a_b
    \,.
\end{equation}
These identities ensure completeness and orthogonality for the symmetric (triplet) subspace of rank-two $\mathrm{SU}(2)$ tensors.
For the antisymmetric (singlet) subspace,
\begin{equation}
    (\sigma_0)^{IJ} (\sigma^0)_{KL} = -2\delta^I_{[K} \delta^J_{L]}
    \,,\quad
    (\sigma^0)_{IJ} (\sigma_0)^{IJ}  = -2
    \,,
\end{equation}
where the Clebsch-Gordan coefficients are given by $(\sigma^0)_{IJ} = -\epsilon_{IJ}$ and $(\sigma_0)^{IJ} = -\epsilon^{IJ}$.
Then, the matrices $(\sigma_0)_I{}^J$ and $(\sigma_a)_I{}^J$ together serve as the Hermitian generators of $\mathrm{U}(2)$:
\begin{align}
\sigma^A = (I, \sigma^a)
\,,\quad
\sigma_A = (-I, \sigma_a) = \eta_{AB} \s^B
\,.
\end{align}
Finally, the singlet-triplet decomposition of a rank-two $\mathrm{SU}(2)$ tensor is defined as
\begin{equation}
    \label{eq:SU2-A-to-IJ}
    X_{IJ} = (\s^A)_{IJ} X_A
    \,,\quad
    X_0 = -\frac{1}{2} (\s_0)^{IJ} X_{IJ} = \frac{1}{2} X_I{}^I
    \,,\quad
    X_a = -\frac{1}{2} (\s_a)^{IJ} X_{IJ}
    \,.
\end{equation}
For complex $X_A$, the behavior under complex conjugation is
\begin{equation}
    [(\sigma_A)_I{}^J]^* = (\sigma_A)_J{}^I
    \,\,\implies\,\,
    [(\sigma^A)_{IJ}]^* = -(\sigma^A)^{JI}
    \,,\quad
    [X_{IJ}]^* = -\bar{X}^A (\s_A)^{JI} := -\bar{X}^{JI} 
    \,.
\end{equation}

\paragraph{$\mathrm{SL}(2,\mathbb{C})$ spinors} 
Our convention is such that the epsilon tensors of different types coincide numerically: 
\begin{align}
    \epsilon^{\a\b} 
    \deq \bar{\epsilon}^{\da\db} 
    \deq  \epsilon^{IJ} 
    \,,\quad
    \epsilon_{\a\b} 
    \deq \bar{\epsilon}_{\da\db} 
    \deq  \epsilon_{IJ} 
    \,.
\end{align}
Then, the rules for raising and lowering $\mathrm{SL}(2,\mathbb{C})$ spinor indices ($\a,\b,\cdots$ and $\da,\db,\cdots$) are the same as that of $\mathrm{SU}(2)$ spinor indices.

We work in the mostly-plus metric signature $\eta_{\m\n} = \mathrm{diag}(-1,+1,+1,+1)$ 
and use the widely-adopted convention for gamma matrices: 
\begin{align}
\gamma^\m =
\begin{pmatrix}
0 & (\sigma^\m)_{\alpha\dot{\beta}}
\\
(\bar{\sigma}^\m)^{\dot{\alpha}\beta} & 0 
\end{pmatrix} \,, 
\quad 
\sigma^\m =  (I,\vec{\s}) \,,
\quad 
\bar{\sigma}^\m =  (I, -\vec{\s}) \,.
\end{align}
The Clifford algebra takes the form $\{\gamma^{\m},\gamma^{\n}\} = -2\eta^{\m\n}$.
Note also that
\begin{gather}
    (\bar{\sigma}^{\m})^{\da\a} = \bar{\epsilon}^{\da\db} \epsilon^{\a\b} \sigma^{\m}_{\b\db}
    \,,
    \quad
    \eta_{\m\n} \s^{\m}_{\a\da} \s^{\n}_{\b\db} 
    = -2\mem\e_{\a\b} \bar{\e}_{\da\db}
    \,.
\end{gather}

Vectors are converted to mixed bi-spinors and vice versa by the following rule:
\begin{equation}
    V_{\alpha\dot{\alpha}} = (\sigma^{\m})_{\alpha\dot{\alpha}} V_{\m} 
    \,,\quad
    V_{\m} = - \frac{1}{2} (\bar{\sigma}_{\m})^{\dot{\alpha}\alpha} V_{\alpha\dot{\alpha}}
    \,.
    \label{eq:vector2spinor}
\end{equation}
For anti-symmetric rank 2 tensors, we choose a particular convention, 
\begin{align}
    \frac{1}{2}\gamma^{[\m}\gamma^{\n]} = 
    \begin{pmatrix}
    (-\s  ^{\m\n})_\alpha{}^\beta & 0 
    \\
    0 & (\bar{\s  }^{\m\n})^{\dot{\alpha}}{}_{\dot{\beta}} &
    \end{pmatrix} \,, 
    \quad 
    (\s  ^{\m\n}) = -\frac{1}{2}\sigma^{[\m} \bar{\sigma}^{\n]} \,,
    \quad 
    (\bar{\s  }^{\m\n}) = \frac{1}{2} \bar{\sigma}^{[\m} \sigma^{\n]} 
    \,,
\end{align}
while $\frac{i}{2}\gamma_{[\m}\gamma_{\n]}$ 
are the Hermitian generators of Lorentz symmetry.
Contracted with the epsilon tensors, the $(2\times 2)$ blocks of $\frac{1}{2}\gamma^{[\m}\gamma^{\n]}$ give symmetric matrices in spinor indices:
\begin{align}
    (\s  ^{\m\n})_{\a\b} := \e_{\b\g} (\s  ^{\m\n})_\a{}^\g  \,,
    \,\,
    (\bar{\s  }^{\m\n})_{\da\db} := \bar{\e}_{\da\dg} (\bar{\s  }^{\m\n})^{\dg}{}_\db 
    \;\;\implies\;\;
    [(\s  ^{\m\n})_{\a\b}]^* = (\bar{\s  }^{\m\n})_{\da\db} \,.
\end{align}
An anti-symmetric tensor is converted to unmixed bi-spinors as 
\begin{align}
    F_{\a\b} \, \bar{\e}_{\da\db} + \e_{\a\b}\, \bar{F}_{\da\db} := 
    \frac{1}{2} F_{\m\n} \left[ (\s  ^{\m\n})_{\a\b} \, \bar{\e}_{\da\db} + \e_{\a\b}\, (\bar{\s  }^{\mu\nu})_{\da\db} \right] 
    = 
    \frac{1}{2} F_{\m\n} \s^{[\m}_{\a\da} \s^{\n]}_{\b\db} 
    \,.
    \label{eq:tensor2spinor}
\end{align}
The spinor notation splits the self-dual and anti-self-dual parts of 
an anti-symmetric tensor:
\begin{align}
\frac{1}{2} \ve^{\m\n}{}_{\r\s} \bar{\s  }^{\r\s} = +i \bar{\s  }^{\m\n} \,, 
\quad
\frac{1}{2} \ve^{\m\n}{}_{\r\s} \s  ^{\r\s} = -i \s  ^{\m\n} \,,
\quad 
\ve_{0123} = +1 \,.
\end{align}

\paragraph{Massless twistors} 

The massless twistor space consists of four complex variables. 
It transforms in the fundamental representation of an $\mathrm{SU}(2,2)$ group, 
to be identified with the $\mathrm{SO}(2,4)$ conformal group of the Minkowski space.
Luckily, the twistor space admits a symplectic structure, opening the possibility 
of interpreting it as the phase space of a massless spinning particle.
We follow a physicist's convention for twistor variables similar (but not identical) to the one of \cite{Witten:2003nn}.
We denote the twistor coordinates by $Z_\mathrm{A}$, their complex conjugates by $\bar{Z}_{\bar{\mathrm{A}}}$, 
and the $\mathrm{SU}(2,2)$ action by 
$Z_{\mathrm{A}} \mapsto (g)_{\mathrm{A}}{}^{\mathrm{B}} Z_{\mathrm{B}}$. 
The indefinite signature of $\mathrm{SU}(2,2)$ is encoded in a ``metric'' $A^{\bar{\mathrm{A}}\mathrm{B}}$:
\begin{align}
    (g^\dagger)^{\bar{\mathrm{A}}}{}_{\bar{\mathrm{C}}} A^{\bar{\mathrm{C}}\mathrm{D}} 
    (g)_{\mathrm{D}}{}^{\mathrm{B}} = A^{\bar{\mathrm{A}}\mathrm{B}} \,.
\end{align}
The metric defines the invariant ``norm'' of the twistor:
\begin{align}
    \bar{Z}^{\mathrm{B}} Z_{\mathrm{B}} := A^{\bar{\mathrm{A}}\mathrm{B}} \bar{Z}_{\bar{\mathrm{A}}} Z_{\mathrm{B}}  \,.
\end{align}
In relating the twistor space to the Minkowski space,  
it is customary to decompose $Z_{\mathrm{A}}$ into two $\mathrm{SL}(2,\mathbb{C})$ spinors
such that
\begin{equation}
    Z_{\mathrm{A}}
    = \begin{pmatrix}
        \lambda_{\alpha}
        \\
        i \mu^{\dot{\alpha}}
    \end{pmatrix}
    \,,
    \quad 
    A^{\bar{\mathrm{A}}\mathrm{B}} 
    = \begin{pmatrix}
        0 & \delta^{\da}{}_{\db} 
        \\
        \delta_{\a}{}^{\b}  & 0 
    \end{pmatrix} \,, 
    \quad 
    \bar{Z}^B 
    = \begin{pmatrix}
        -i \bar{\mu}^\b & \bar{\l}_\db  
    \end{pmatrix} \,. 
    \label{eq:Z-A-Zbar}
\end{equation}
The factors of $(\pm i)$ in front of $\mu$ and $\bar{\mu}$ have been inserted for later convenience. 

The $\mathrm{SU}(2,2)$-invariant symplectic structure defined on the twistor space is
\begin{align}
    \omega
    &=  i (d \bar{Z}^{\mathrm{A}} \wedge d Z_{\mathrm{A}})
    =  d\bar{\mu}^{\alpha} \wedge d\lambda_{\alpha} + d \mu^{\dot{\alpha}} \wedge d \bar{\lambda}_{\dot{\alpha}} 
    \,.
\end{align}
The non-vanishing Poisson brackets are
\begin{align}
    \label{eq:massless-twistor-PB}
    \{ Z_{\mathrm{A}}, \bar{Z}^{\mathrm{B}} \} = - i\mem \delta_\mathrm{A}{}^\mathrm{B}
    \quad\implies\quad
    \{\lambda_{\alpha},\bar{\mu}^{\beta}\} =  \delta_{\alpha}{}^{\beta}
    \,,\quad
    \{\bar{\lambda}_{\dot{\alpha}}, \mu^{\dot{\beta}}\} =  \delta_{\dot{\alpha}}{}^{\dot{\beta}}
    \,.
\end{align}
The Lie algebra $\mathfrak{su}(2,2)$ manifests itself as a set of Hamiltonian flows satisfying $dG = -\mathrm{i}_V \omega$, 
where $G$ is a function and $V$ is a vector field. A convenient choice of basis for $G$ is 
\begin{align}
 G_{\mathrm{A}}{}^{\mathrm{B}} := 
 Z_{\mathrm{A}} \bar{Z}^{\mathrm{B}}  - \frac{1}{4}  \delta_\mathrm{A}{}^\mathrm{B} (\bar{Z}^{\mathrm{C}} Z_{\mathrm{C}})\,.   
    \label{eq:su22-gen}
\end{align}
Their Poisson brackets are
\begin{align}
    \{ G_{\mathrm{A}}{}^{\mathrm{B}} , G_{\mathrm{C}}{}^{\mathrm{D}} \} 
    = - i (\delta_\mathrm{A}{}^\mathrm{D}  G_{\mathrm{C}}{}^{\mathrm{B}} -  \delta_\mathrm{C}{}^\mathrm{B}  G_{\mathrm{A}}{}^{\mathrm{D}} ) \,.
\label{su22}
\end{align}

To connect the twistor space to the Minkowski space further, 
we rewrite the conformal generators of the latter in the spinor notation using \eqref{eq:vector2spinor} 
with one and only exception: for position variables,
\begin{equation}
    x^{\da\a} := -\frac{1}{2} (\bs_\m)^{\da\a} x^\m
    \,,\quad
    x^\mu = (\s^\mu)_{\a\da} x^{\da\a}
\end{equation}
such that $\partial/\partial x^{\da\a} = (\s^\m)_{\a\da} \partial/\partial x^\m$.
This rescaling of $x^{\da\a}$ reduces unwarranted factors of $2$ in many equations, 
starting from the fundamental Poisson bracket in the Minkowski space:
\begin{align}
    \{ x^\m, p_\n \} = \delta^\m{}_\n 
    \quad 
    \implies
    \quad 
    \{ x^{\da\a}, p_{\b\db}\} = \dt^\a_\b \dt^\da_\db 
    \,.
\end{align}
The conformal algebra boils down to
\begin{align}
    \begin{split}
        \{ J_{\a\b} , J_{\g\delta} \} &=  - \frac{1}{2} \left( \e_{\a\g} J_{\b\delta} + \e_{\b\g} J_{\a\delta} + \e_{\a\delta} J_{\b\g} + \e_{\b\delta} J_{\a\g} \right)  \,, 
        \\
        \{ J_{\a\b}, P_{\g\dg} \} &=  - \frac{1}{2} \left( \e_{\a\g} P_{\b\dg} + \e_{\b\g} P_{\a\dg} \right) \,, \quad\,\, \{ D , P_{\a\da} \} = - P_{\a\da} \,,
        \\
        \{ J_{\a\b}, K_{\g\dg} \} &= - \frac{1}{2} \left( \e_{\a\g} K_{\b\dg} + \e_{\b\g} K_{\a\dg} \right) \,, \quad \{ D , K_{\a\da} \} = + K_{\a\da} \,,
        \\
        \{ P_{\a\da}, K_{\b\db} \} & = - \left( \e_{\a\b} \bar{J}_{\da\db} + \bar{\e}_{\da\db} J_{\a\b} - \e_{\a\b}\bar{\e}_{\da\db} D \right) \,. 
    \end{split}
    \label{conf-spin2}
\end{align}
The Poincar\'e generators $P_\m = p_\m$ and $J_{\m\n} = x_{\m}p_{\n} - x_{\n}p_{\m}$ constitute the conformal algebra 
together with the dilatation generator $D = - x^\m p_\m$ and the special conformal generator $K^{\da\a} = x^{\da\b} p_{\b\db} x^{\db \a}$.
The normalization of $K^{\da\a}=(\bs_\m)^{\da\a}K^\m$ is such that $K_{\m} = IP_{\m}I$ holds with the inversion map $I: -x^\m/2 \mapsto -2\,{x^\m/x^2}$.

Now, we identify the $\mathfrak{su}(2,2)$ generators in \eqref{eq:su22-gen} with the conformal generators as 
\begin{align}
    \begin{split}
        G_A{}^B  =  
        \begin{pmatrix}
            -i J_\a{}^\b - \textstyle{\frac{i}{2}} \delta_\a{}^\b D  & - P_{\a\db} 
            \\ 
            - K^{\da\b} &  +i \bar{J}^\da{}_{\db}  + \textstyle{\frac{i}{2}} \delta^\da{}_{\db} D  
        \end{pmatrix} \,.
    \end{split}
    \label{eq:su22-conf-map}
\end{align}
More explicitly,  
\begin{align}
    \begin{split}
        J_{\a\b} =  \lambda_{(\a} \bar{\mu}_{\b)}\,, 
        \quad
        &D = \frac{1}{2} \left( \bar{\mu}^\a \lambda_\a +  \bar{\lambda}_\da \mu^\da  \right) \,,
        \\
        P_{\a\da} = - \lambda_\a \bar{\lambda}_\da \,,
        \quad 
        &K^{\da\a} = - \mu^\da \bar{\mu}^\a  \,.
    \end{split}
\end{align}
It is straightforward to verify that \eqref{su22} and \eqref{eq:su22-conf-map} reproduce \eqref{conf-spin2}.

We finally turn to the (complexified) incidence relation, 
\begin{equation}
    \mu^\da =  
    z^{\dot{\alpha}\beta}
     \lambda_{\beta} \,.
    \label{eq:incidence-massless}
\end{equation}
Substituting $z^{\dot{\alpha}\beta} {\lambda}_{\beta}$ for $\mu^\da$
into  $\bar{J}_{\dot{\alpha}\dot{\beta}} = \bar{\lambda}_{(\da} \mu_{\db)}$ 
and using $p_{\a\da} = - \lambda_\a \bar{\lambda}_\da$, we find
\begin{gather}
    \bar{J}_{\dot{\alpha}\dot{\beta}} 
    = \bar{\lambda}_{(\dot{\alpha}\vphantom{\db}} z_{\dot{\beta})}{}^{\gamma} \lambda_{\gamma}
    = z_{\gamma(\dot{\alpha}\vphantom{\db}} p_{\dot{\beta})}^{\gamma}
    \,.
    \label{eq:Jzp}
\end{gather}
In the vector notation with $z^\m = x^\m + i y^\m$, it means $J_{\m\n} = x_\m p_\n - x_\n p_\m + *(y_\m p_\n - y_\n p_\m)$, where 
$*$ denotes the Hodge dual. 
The real part of $z^\mu$ clearly parametrizes the position in the Minkowski space.  
The imaginary part can be used to account for the helicity of a massless particle. 
One way to see it is that when $\mathfrak{sl}(2,\mathbb{C})$ is regarded as the complexification of $\mathfrak{su}(2)$,  
the orbital and spin angular momenta take the real and imaginary parts, respectively. 
To see it in another way, we split the real and imaginary parts of 
\begin{align}
    \begin{split}
        -z^{\da \a} p_{\a \da} &= \bar{\lambda}_\da z^{\da \a} \lambda_\a = \bar{\lambda}_\da \rmu^\da \,,
        \\
        -\bar{z}^{\da\a} p_{\a \da} &= \bar{\lambda}_\da \bar{z}^{\da\a} \lambda_\a = \bar{\mu}^\a \lambda_\a \,,
    \end{split}
\end{align}
to find
\begin{align}
    \begin{split}
    -x^{\a\da} p_{\da \a} &= \frac{1}{2} \left(
        \bar{\lambda}_\da \rmu^\da
        +
        \lmu^\a \lambda_\a
    \right)
    = D
    \,,
    \\
    -y^{\da \a} p_{\a \da} &= \frac{1}{2i} \left(
        \bar{\lambda}_\da \rmu^\da
        -
        \lmu^\a \lambda_\a
    \right) 
    = -\frac{1}{2} \bar{Z}^{\mathrm{A}} Z_{\mathrm{A}}
    \,.
    \end{split}
\end{align}
The first line is consistent with the map \eqref{eq:su22-conf-map}. 
The second line relates the helicity with the generator of the overall U$(1)$ commuting with SU$(2,2)$. 
The same generator appears as the proportionality constant between the Pauli-Lubanski vector 
$-(*J)_{\m\n} P^\n$ and the momentum $P_\m$. 
Interpreting this U$(1)$ as the helicity of a massless particle also plays a crucial role in the modern application of 
twistor theory to scattering amplitudes.


\end{document}